\documentclass[12pt]{article}
\usepackage{amsmath}
\usepackage{amssymb}
\usepackage{amsbsy}
\renewcommand{\theequation}{\arabic{section}.\arabic{equation}}
\newcommand{\newsec}{\setcounter{equation}{0} \section}
\newcommand{\appsec}{\setcounter{section}{0} \setcounter{equation}{0}
\renewcommand{\theequation}{\Alph{section}.\arabic{equation}}
\renewcommand{\thesection}{\Alph{section}} \section} \newcounter{temp}
\newenvironment{subequation}{
\setcounter{temp}{\value{equation}}
\addtocounter{temp}{1}
\setcounter{equation}{0}
\renewcommand{\theequation}
{\arabic{section}.\arabic{temp}\alph{equation}} }%
{\setcounter{equation}{\value{temp}}
\renewcommand{\theequation}{\arabic{section}.\arabic{equation}} }
\newenvironment{asubequation}{
\setcounter{temp}{\value{equation}}
\addtocounter{temp}{1}
\setcounter{equation}{0}
\renewcommand{\theequation}
{\Alph{section}.\arabic{temp}\alph{equation}}}%
{\setcounter{equation}{\value{temp}}
\renewcommand{\theequation}{\Alph{section}.\arabic{equation}} }

\newtheorem{lemma}{Lemma}

\begin{document}
\bibliographystyle{siam}
\title{Global homogenization of a dilute suspension of spheres.
Suspension
rheology}
\author{Y. ALMOG \thanks{Department of Mathematics, M.I.T, Cambridge, MA
02139-4307,
USA.} \hspace{0em} \footnotesize AND \large \hspace{0em} H. BRENNER
\thanks{Department of Chemical Engineering, M.I.T, Cambridge, MA
02139-4307, USA.}}
\date{}
\maketitle
\abstract
{A new method for rheologically homogenizing a dilute suspension
composed
of freely-suspended spherical particles dispersed in a Newtonian fluid
is
presented: The ensemble-averaged velocity and stress fields obtained for
the neutrally-buoyant sphere suspension are compared with the respective
velocity and stress fields
obtained for a hypothetical homogeneous Newtonian fluid continuum
possessing a spatially {\em
non-uniform} viscosity for the
same specified boundaries and ambient flow. The method is global in
nature;
that is, wall effects and spatial dependence of both the ambient flow
and
the particle number density are encountered, thereby confirming known
classical results up to $O(c^{2})$ terms ($c$ = volume concentration of
spheres) for
the suspension viscosity which have previously been obtained by assuming
{\em a priori} that the suspension is both unbounded and statistically
homogeneous.}
\section {Introduction}
The rheology of a dilute suspension composed of freely-suspended rigid
spherical particles in a Newtonian fluid has been extensively studied
since
Einstein's \cite{ei06} $O(c)$ classical analysis. His calculation
related
the
increased viscosity of the suspension to the additional dissipation
occurring within a 'suspension cell' owing to the perturbing presence of
a
freely suspended sphere in an otherwise uniform shear field. This
classical
rheological result is
supported by the analysis of Keller {\em et al.} \cite{keetal67}, who
used
variational principles to bound the overall dissipation in a suspension
dispersed in a homogeneous shear field under the assumption that most
particles lie relatively far from one other.

Such scalar dissipation arguments are viable only in cases where the
suspension behaves macroscopically as a homogeneous isotropic fluid. In
particular, these methods are inapplicable in circumstances where the
suspension-scale stress/rate-of-strain relationship is anisotropic.
Batchelor \cite{ba70} and Brenner
\cite{br72}, building on the pioneering work of Kirkwood \cite{ki67a},
\cite{ki67b} and Giesekus
\cite{gi64}, developed a general theory from which the
stress/strain-rate
relation may be obtained. Their methods are based on calculations of the
average interstitial-scale stress and velocity gradient tensors, such
averaging being performed over a 'suspension cell.' In the case of
freely
suspended spherical particles an isotropic rheological constitutive
relation is obtained.

Higher-order terms in the relative-viscosity/suspended-particle
concentration expansion have been obtained by Batchelor \& Green
\cite{bagr72a},\cite{bagr72b}.
Their method is based on an 'ensemble-average' approach; that is, they
obtain the relation between the averages over all possible $N$-sphere
configurations of the stress and rate-of-strain tensors. To obtain
$O(c^{2})$ terms in this expansion, only two-sphere configurations need
be
considered (a point
to which we will subsequently return). Cox \& Brenner \cite{cobr71}
developed an
alternative scheme to obtain these $O(c^{2})$ terms, although their
generic
methods have not yet been implemented in the context of a specific
rheological problem.

Each of the previously cited methods is essentially local in nature;
that
is, effects of bounding walls as well as spatial inhomogeneities in the
ambient velocity gradient are neglected. When the ensemble-average
approach
is applied, and the existence of walls ignored, nonconvergent integrals
arise (presumably owing to the non-uniformly valid nature of the
Stokes-flow approximation in infinite domains). To overcome this
difficulty, {\em ad hoc} renormalization methods
\cite{bagr72a},\cite{bagr72b}, based largely on intuitive arguments,
have
been invoked.

Hinch \cite{hi77} developed another renormalization technique (his
so-called ``second renormalization'') which enabled the calculation of
the
permeability of a random fixed bed of spherical particles. He also
offered
further physical insights into the nature of Batchelor \& Green's
\cite{bagr72b} renormalization scheme, but did not formally resolve the
underlying issues. A different insight into renormalization methods was
later offered by O'Brien \cite{ob79}, who effectively removed the
convergence problem at infinity by correctly adding a ``macroscopic
boundary integral''. Both Hinch's \cite{hi77} and O'Brien's \cite{ob79}
analyses are again local in nature in the sense that the domain
investigated is assumed {\em a priori} to be unbounded as well as
statistically homogeneous.

The subsequent analysis develops an ensemble-average technique, via
which
we obtain the suspension's
average velocity and stress fields. Wall effects and spatial dependence
of
the ambient flow are encountered, and renormalization is not needed.
Results expressed in the form of integral representations, in which the
kernel is Green's function and the density function is the surface
traction, are derived and compared with the respective velocity and
stress
fields obtained for a hypothetical homogeneous fluid possessing a {\em
non-uniform} viscosity, for the same specified boundaries and ambient
flow.
It is demonstrated that for a certain (unique) choice of spatially {\em
non-uniform} viscosity field, the velocity and stress fields obtained
for
the
hypothetical homogeneous medium are asymptotically equal to the
respective
averages obtained for the suspension throughout the whole domain, except
for a thin boundary layer near the walls. Though the results for the
suspension-average velocity and stress fields depend upon the boundary's
size and shape, the results for the viscosity field of the hypothetical
homogeneous medium are domain independent. The comparison is made up to
$O(c^{2})$ terms, and the known results of Einstein \cite{ei06} and
Batchelor \& Green \cite{bagr72a},\cite{bagr72b} formally confirmed.

Consider $N$ identical rigid spherical particles freely suspended in a
homogeneous Newtonian fluid of viscosity $\mu$. The sphere centers are
respectively situated at the points ($\boldsymbol{x}_{1},\ldots,
\boldsymbol{x}_{N})$. (For later reference we shall term such a set of
locations a `configuration'.) Denote by $\Omega$ the domain, and by
$(\bar{\boldsymbol{u}},\bar{p})$ the ambient flow which satisfies the
boundary conditions on $\partial \Omega$. Two different types of
boundary
conditions will be considered. On one subset of the boundary, denoted by
$\partial \Omega_{u}$, we prescribe the velocity ('adherence-to-walls'),
whereas on the complementary subset, denoted by $\partial \Omega_{f}$
($\partial \Omega = \partial \Omega_{u} \cup \partial \Omega_{f}$), we
prescribe the surface traction.

We shall confine the discussion to finite domains
(diam($\Omega)=R^{\prime}$) or though the analysis can be extended, in
principle, to
domains which are infinite in one direction, e.g., a cylinder. In the
latter case, $R^{\prime}$ is determined by the cross section of the
cylinder. We assume that the ambient flow varies on a length scale of
$O(R^{\prime})$. The length scale characterizing the number density
$n(\boldsymbol{x})$ is assumed to be of the same order. This number
density
is defined as the number of particles per unit volume in any domain of
characteristic size $L$ which is much smaller than $R^{\prime}$ but much
larger than the average distance $l=(N/V)^{-1/3}$ between neighboring
particles, where $N$ is the total number of spheres in $\Omega$, and $V$
is
the domain's volume. (In the case of a cylinder, $N/V$ should be
interpreted as the number density of particles per unit length.)

It is convenient to non-dimensionalize the spatial coordinate by $l$.
Denote the dimensionless radius of the suspended spheres by $\epsilon$
($\epsilon \ll1$), supposed small since the suspension is assumed
dilute.
The dimensionless
length
scale $R^{\prime}/l$ of the ambient flow will be denoted by $R$ ($R\gg
1$).

When inertial effects are negligible the velocity and pressure fields
may
be derived from the respective integral representations \cite{liba92}:
\begin{equation}
\label{int10}
u_{i}(\boldsymbol{y},\boldsymbol{x}_{1},\ldots,\boldsymbol{x}_{N}) =
\bar{u}_{i}(\boldsymbol{y}) +
\sum_{n=1}^{N} \int_{\partial s_{n}}
T_{ij}(\boldsymbol{x},\boldsymbol{y})
f_{j}(\boldsymbol{x}) ds_{x},
\end{equation}

\begin{equation}
\label{int20}
p(\boldsymbol{y},\boldsymbol{x}_{1}, \ldots ,\boldsymbol{x}_{N}) =
\bar{p}(\boldsymbol{y}) + \sum_{n=1}^{N} \int_{\partial s_{n}}
P_{j}(\boldsymbol{x},\boldsymbol{y})
f_{j}(\boldsymbol{x}) ds_{x},
\end{equation}
wherein \(\boldsymbol{u}\) and \(p\) are the configuration-dependent
velocity and pressure fields, respectively. The vector
\(\boldsymbol{f}\)
is the surface traction, \begin{equation}
\boldsymbol{f} = \boldsymbol{\sigma(u)}\cdot\hat{\boldsymbol{n}},
\end{equation}
where \(\boldsymbol{\sigma (u)}\) is the stress tensor deriving from
\((\boldsymbol{u},p)\), and $\hat{\boldsymbol{n}}$ is the inward unit
normal; \( {\partial s_{n}}\) denotes the surface (\(
|\boldsymbol{x}-\boldsymbol{x}_{n}|= \epsilon \)) of the n`th sphere,
and $\boldsymbol{T}(\boldsymbol{x},\boldsymbol{y})$,
\(\boldsymbol{P}(\boldsymbol{x},\boldsymbol{y})\) are the Green's
functions
respectively defined by
\begin{eqnarray}
T_{ij} = t_{ij} + \tau_{ij} & , & P_{i} = p_{i} + \pi_{i}.
\end{eqnarray}
In the latter, \((\boldsymbol{t,p})\) denotes the Stokeslet
\begin{equation}
\left.
\begin{array}{ccc}
t_{ij}(\boldsymbol{x,y}) & = & \displaystyle{\frac{1}{8 \pi \mu}
\left(\frac{\delta_{ij}}{r} + \frac{r_{i} r_{j}}{r^3} \right)}
\\
& & \\
p_{i}(\boldsymbol{x,y}) & = & \displaystyle{\frac{1}{4 \pi}
\frac{r_i}{r^{3}}}
\end{array}
\right\} \: \boldsymbol{r=x-y} \; ,
\end{equation}
with \((\boldsymbol{\tau ,\pi})\) the Stokeslet image, the latter being
a
regular solution of the Stokes problem satisfying the boundary condition
$\left. \tau_{ij}\right|_{\partial \Omega_{u}} = -\left. t_{ij}
\right|_{\partial \Omega_{u}}$ , $\left. \sigma_{ij}({\boldsymbol
\tau}_{.k}) \cdot \hat{n}_{j} \right|_{\partial \Omega_{f}} = -\left.
\sigma_{ij}({\boldsymbol
t}_{.k}) \cdot \hat{n}_{j} \right|_{\partial \Omega_{f}}$, where
${\boldsymbol t}_{.k}$ denotes the vector $(t_{1k},t_{2k},t_{3k})$. The
representation (\ref{int10}) and (\ref{int20}) [as well as
(\ref{int15})]
are still valid for cylindrical domains, where $N$ can be infinite. For
instance, since all particles are neutrally buoyant it is easy to show
(cf. also appendix A) that as $|\boldsymbol{x}_{n} - \boldsymbol{y}|
\to \infty$,
\begin{displaymath}
\int_{\partial s_{n}}
T_{ij}(\boldsymbol{x},\boldsymbol{y})
f_{j}(\boldsymbol{x}) ds_{x} \sim O(|\boldsymbol{x}_{n} -
\boldsymbol{y}|^{-2}) \; .
\end{displaymath}
If the particles are appropriately numbered, then $|\boldsymbol{x}_{n} -
\boldsymbol{y}| \geq C(\epsilon,\Omega)n$, whence the series appearing
in (1.1)
is absolutely convergent.

It is easy to show using (\ref{int10}) and (\ref{int20}) that the stress
field may be expressed in the form 
\begin{equation}
\label{int15}
\sigma_{ij}(\boldsymbol{y},\boldsymbol{x}_{1},\ldots,\boldsymbol{x}_{N})
=
\bar{\sigma}_{ij}(\boldsymbol{y}) + \sum_{n=1}^{N} \int_{\partial
s_{n}} (\sigma_{y})_{ij}(\boldsymbol{T}_{.k} (\boldsymbol{x,y}))
f_{k}(\boldsymbol{x,x}_{1},\ldots,\boldsymbol{x}_{N}) ds_{x}.
\end{equation}
Upon letting $\boldsymbol{y}$ approach the surface of one of the
particles,
dot-multiplying by the inward normal, and using the 'jump condition'
\cite{la63}, it may be shown \cite{liba92} that the surface traction
$\boldsymbol{f}$ satisfies the
boundary integral equation
\begin{equation}
\label{int30}
\frac{1}{2}
f_{i}(\boldsymbol{y},\boldsymbol{x}_{1},\ldots,\boldsymbol{x}_{N}) =
\bar{f}_{i}(\boldsymbol{y}) +
\sum_{n=1}^{N} \int_{\partial s_{n}}
(\sigma_{y})_{ij}(\boldsymbol{T}_{.k}
(\boldsymbol{x,y}))
f_{k}(\boldsymbol{x,x}_{1},\ldots,\boldsymbol{x}_{N})
ds_{x} \hat{n}_{j}(\boldsymbol{y})
\end{equation}
for \(\boldsymbol{y} \in \bigcup_{n=1}^{N}\partial s_{n}\), wherein
$\bar{\boldsymbol{f}}$ is the surface traction due to
$(\bar{\boldsymbol{u}},\bar{p})$, and \(
\boldsymbol{\sigma}_{y}(\boldsymbol{T}_{.k}) \) is the stress tensor due
to
$(\boldsymbol{T}_{.k},\boldsymbol{P}_{k})$.

In the next section we introduce an iterative scheme used to approximate
the solution of (\ref{int30}), thereby obtaining the average velocity
and
stress fields. Section 3 demonstrates the equivalence of the suspension
with the homogeneous medium. Section 4 addresses several key points,
insufficiently emphasized within the analysis. In
appendix A we prove, upon invoking some mild assumptions, that the
scheme
presented in Section 2 is asymptotically accurate. Appendix B shows that
the effect on the surface traction of any walls bounding the flow, as
well
as any spatial dependence of the ambient flow, has a negligible effect
upon the average velocity field, thereby justifying the analysis of
Section 3.

\newsec{The average velocity field}
We begin by introducing the iterative scheme by which we approximate the
solution of (\ref{int30}) (or the surface traction field). The first
iteration is obtained by solving the one-sphere problem,
\begin{equation}
\label{av10}
\frac{1}{2} f_{0i}^{(n)}(\boldsymbol{y},\boldsymbol{x}_{n}) =
\bar{f}_{i}(\boldsymbol{y}) +
\int_{\partial s_{n}} (\sigma_{y})_{ij}(\boldsymbol{T}_{.k}
(\boldsymbol{x,y})) f_{0k}^{(n)}(\boldsymbol{x},\boldsymbol{x}_{n})
ds_{x}\hat{n}_{j}(\boldsymbol{y}),
\end{equation}
for \(\boldsymbol{y} \in \partial s_{n} \). The superscript $(n)$
denotes the
fact that unlike $\boldsymbol{f}$, the distribution
$\boldsymbol{f}^{(n)}$
is defined
only over
\( \partial s_{n} \).
The second iteration is the solution of yet another one-sphere problem,
namely
\begin{eqnarray}
\label{av20}
\frac{1}{2}
f_{1i}^{(n_{1})}(\boldsymbol{y},\boldsymbol{x}_{1},
\ldots,\boldsymbol{x}_{N})
= \bar{f}_{i}(\boldsymbol{y}) +
\int_{\partial s_{n}} (\sigma_{y})_{ij}(\boldsymbol{T}_{.k})
f_{1k}^{(n_{1})}(\boldsymbol{x,x}_{1},\ldots,\boldsymbol{x}_{N})
ds_{x}\hat{n}_{j}(\boldsymbol{y}) \nonumber \\ + \sum_{\stackrel
{\scriptstyle n_{2} = 1} {n_{2} \neq n_{1}}} ^{N} \int_{\partial s_{n}}
(\sigma_{y})_{ij}(\boldsymbol{T}_{.k})
f_{0k}^{(n_{1},n_{2})}(\boldsymbol{
x,x}_{n_{1}},\boldsymbol{x}_{n_{2}}) ds_{x} \hat{n}_{j}(\boldsymbol{y}).
\end{eqnarray}
Herein, \( \boldsymbol{f}_{0}^{(n_{1},n_{2})} \) is a solution of the
following
two-sphere problem:
\begin{eqnarray}
\label{av30}
\frac{1}{2}
f_{0i}^{(n_{1},n_{2})}(\boldsymbol{y,x}_{n_{1}},\boldsymbol{x}_{n_{2}})
=
\bar{f}_{i}(\boldsymbol{y}) + \int_{\partial
s_{n_1}\bigcup \partial
s_{n_2}}(\sigma_{y})_{ij}(\boldsymbol{T}_{.k})
f_{0k}^{(n_{1},n_{2})}(\boldsymbol{x,x}_{n_{1}},\boldsymbol{x}_{n_{2}})
ds_{x} \hat{n}_{j}(\boldsymbol{y}), \\
\boldsymbol{y} \in \partial s_{n_1} \bigcup \partial s_{n_2}). \nonumber
\end{eqnarray}
Equation (\ref{av20}) includes the effect of touching, two-sphere
hydrodynamic interactions. The ideas underlying (\ref{av20}) closely
resemble those employed in the method of scattering by groups
\cite{ruke89b}. Note that \hspace {0.3cm} \(
\boldsymbol{f}_{0}^{(n_{1},n_{2})}
(\boldsymbol{x,x}_{n_{1}},\boldsymbol{x}_{n_{2}}) \sim
\boldsymbol{f}_{0}^{(n_{2})} (\boldsymbol{x,x}_{n_{2}}) \) \hspace
{0.3cm}
for
\(\boldsymbol{x} \in \partial s_{n_2} \) when \( \left|
\boldsymbol{x}_{n_1} -
\boldsymbol{x}_{n_2} \right| \gg \epsilon \). Were we to substitute \(
\boldsymbol{f}_{0}^{(n_{2})} \) instead of \(
\boldsymbol{f}_{0}^{(n_{1},n_{2})} \) into (\ref{av20}) (thereby
neglecting
the near-field effect), the resulting approximation for the average
velocity field would be no more accurate than the one based on \(
\boldsymbol{f}_{0} \) (cf. Appendix A). The usual explanation
\cite{bagr72b} underlying the need for addressing two-sphere, near-field
interactions, is that since
the probability of finding a pair of closely proximate spheres is of $O
(c^2)$, errors of $O(1)$ in the configuration-dependent velocity field
arising from the neglect of such near-field interactions cannot be
allowed
when calculating $O(c^2)$ terms in the average velocity field. Some
support
for that intuitive argument is outlined in Appendix A.

It is more convenient to present the traction field in the form of
series
rather than as a successive sequence of iterations. In this context, the
following result is easily obtained:
\begin{eqnarray}
\label{av40}
\boldsymbol{f}_{1}^{(n_{1})}(\boldsymbol{x,x}_{1},\ldots,\boldsymbol{x}_{N})
-
\boldsymbol{f}_{0}^{(n_{1})} (\boldsymbol{ x,x}_{n_{1}}) =
\sum_{\stackrel
{\scriptstyle n_{2} = 1} {n_{2} \neq
n_{1}}}^{N}
\boldsymbol{f}_{0}^{(n_{1},n_{2})}
(\boldsymbol{x,x}_{n_{1}},\boldsymbol{x}_{n_{2}}) -
\boldsymbol{f}_{0}^{(n_{1})} (\boldsymbol{x,x}_{n_{1}}), \\
(\boldsymbol{x}
\in \partial s_{n_1}). \nonumber \end{eqnarray}
If the configuration is kept fixed and $\epsilon$ allowed to tend
towards
zero (and thus $c \to 0$), the right-hand side of the above will
tend
to zero and will explicitly be of $O(\epsilon^3)$ relative to
$\boldsymbol{f}_{0}$, i.e.
\begin{displaymath}
\sum_{\stackrel {\scriptstyle n_{2} = 1} {n_{2} \neq
n_{1}}}^{N} |\boldsymbol{f}_{0}^{(n_{1},n_{2})} -
\boldsymbol{f}_{0}^{(n_{1})}| \leq C \epsilon^3
|\boldsymbol{f}_{0}^{(n_{1})}| \qquad \forall  0<\epsilon<\epsilon_{0}
\;,
\end{displaymath}
for some $\epsilon_0>0$ and $C$ which may depend on the configuration,
the
boundaries etc.

The next step consists of obtaining the configuration-dependent velocity
field. To this end, set
\begin{equation}
\label{av45}
\boldsymbol{u}(\boldsymbol{y,x}_{1},\ldots,\boldsymbol{x}_{N}) =
\bar{\boldsymbol{u}}(\boldsymbol{y}) +
\boldsymbol{u}_{1}(\boldsymbol{y},\boldsymbol{x}_{1},\ldots,\boldsymbol{x}_{N})
+ \boldsymbol{u}_{2}(\boldsymbol{y,x}_{1},\ldots,\boldsymbol{x}_{N}).
\end{equation}
The field \(\boldsymbol{u}_{1}\) is obtained by substituting the
solution of
(\ref{av10}) into (\ref{int10}):
\begin{equation}
\label{av50}
u_{1i}(\boldsymbol{y},\boldsymbol{x}_{1},\ldots,\boldsymbol{x}_{N}) =
\sum_{n=1}^{N} \int_{\partial s_{n}}
T_{ij}(\boldsymbol{x},\boldsymbol{y})
f_{0j}^{(n)}(\boldsymbol{x,x}_{n}) ds_{x}. \end{equation}
Similarly, \(\boldsymbol{u}_{2}\) is obtained by first solving
(\ref{av20}), substituting the result into (\ref{av30}), and
subsequently
introducing the result of the latter operation into (\ref{int10}). This
yields
\begin{equation}
\label{av60}
u_{2i}(\boldsymbol{y},\boldsymbol{x}_{1},\ldots,\boldsymbol{x}_{N}) =
\sum_{\stackrel {\scriptstyle n_{2} = 1} {n_{2} \neq n_{1}}} ^{N}
\int_{\partial s_{n_1}} T_{ij}(\boldsymbol{x},\boldsymbol{y})
[f_{0j}^{(n_{1},n_{2})}(\boldsymbol{x,x}_{n_{1}},\boldsymbol{x}_{n_{2}})
-
f_{0j}^{(n_1)}(\boldsymbol{x,x}_{n_1})] ds_{x}. 
\end{equation}

We next seek to obtain the average velocity field. Denote the
configurational probability density at
$(\boldsymbol{x}_{1},\ldots,\boldsymbol{x}_{N})$ at a specified instant
$t$
by the
multivariable function
$f_{N}(\boldsymbol{x}_{1},\ldots,\boldsymbol{x}_{N},t)$. The average
velocity, \begin{equation}
\label{av70}
\left< \boldsymbol{u} \right>(\boldsymbol{y}) = \int
\boldsymbol{u}(\boldsymbol{y,x}_{1},\ldots,\boldsymbol{x}_{N})f_{N}
(\boldsymbol{x}_{1},\ldots,\boldsymbol{x}_{N},t) dx_{1} \ldots dx_{N}, 
\end{equation}
denoted here by \( \left< \boldsymbol{u} \right> \), is obtained by
averaging the configuration-dependent velocity over all possible
configurations. All other averaged quantities will subsquently be
defined
and designated in the same manner.

Our first goal is that of obtaining \( \left< \boldsymbol{u}_{1} \right>
\). In this context we note that
\begin{equation}
u_{1i} = \sum_{n=1}^{N} u_{1i}^{\prime}(\boldsymbol{x,x}_{n}),
\end{equation}
wherein
\begin{equation}
u_{1i}^{\prime}(\boldsymbol{x,x}_{n}) =\int_{|\boldsymbol{
\xi}|=\epsilon}
T_{ij} (\boldsymbol{x}_{n}+\boldsymbol{\xi},\boldsymbol{x})
f_{0j}^{(n)}(\boldsymbol{ \xi})ds_{\xi}. \end{equation}
Substitution into (\ref{av70}) yields
\begin{equation}
\left< u_{1i} \right> = \sum_{n=1}^{N} \int_{\Omega}
u_{1i}^{\prime}(\boldsymbol{x,x}_{n})
f_{1}^{(n)}(\boldsymbol{x}_{n},t)dx_{n}, \end{equation}
where \(f_{1}^{(n)}(\boldsymbol{x}_{n},t)\) is a first-order marginal
probability density. The quantity
\(f_{1}^{(n)}(\boldsymbol{x}_{n}^{0},t)dx_{n}\) represents the
probability
of finding the n'th particle center $\boldsymbol{x}_{n}$ in the box
\(x_{ni}^0 \leq x_{ni} \leq x_{ni}^0+dx_{ni}\). Due to particle
indistinguishability, it is plausible \cite{ruke89b} that
$f_{N}(\boldsymbol{x}_{1},\ldots,\boldsymbol{x}_{N},t)$ is symmetric
with
respect to all particles. (Otherwise, it may be symmetrized with no
changes
in any of the following averages.) Hence, \begin{displaymath}
f_{1}^{(n)}(\boldsymbol{x}_{n},t)=f_{1}^{(k)}(\boldsymbol{x}_{k},t) \;
\;
\mbox {whenever} \; \; \boldsymbol{x}_{n}=\boldsymbol{x}_{k}.
\end{displaymath}
The number density $n(\boldsymbol{x}_{1})$, which is the probability
density
for
finding any of the $N$ particles at $\boldsymbol{x}_{1}$, may be
obtained as
\begin{displaymath}
n(\boldsymbol{x}_{1}) = Nf_{1}(\boldsymbol{x}_{1}). \end{displaymath}
Consequently,
\begin{equation}
\label{av80}
\left< u_{1i} \right> (\boldsymbol{x}) = \int_{\Omega}
\int_{|\boldsymbol{\xi}|=\epsilon}
T_{ij}(\boldsymbol{x}_{1}+\boldsymbol{\xi},\boldsymbol{x})f_{0j}^{(1)}
(\boldsymbol{ \xi})ds_{\xi} n(\boldsymbol{x}_{1}) dx_{1}. 
\end{equation}
In a similar manner, again taking advantage of the symmetry of
$f_{N}(\boldsymbol{x}_{1},\ldots,\boldsymbol{x}_{N},t)$, we obtain
\begin{equation}
\label{av90}
\left< u_{2i} \right> (\boldsymbol{x})= \int_{\Omega}
n(\boldsymbol{x}_{1})
dx_{1} \int_{\Omega} \int_{|\boldsymbol{\xi}|=\epsilon}
T_{ij}(\boldsymbol{x}_{1}+\boldsymbol{\xi},\boldsymbol{x})[
f_{0j}^{(1,2)}(\boldsymbol{x}_{1},\boldsymbol{x}_{2},\boldsymbol{\xi})
-f_{0j}^{(1)}(\boldsymbol{x}_{1},\boldsymbol{\xi})]ds_{\xi}
P(\boldsymbol{x}_{2}/\boldsymbol{x}_{1}) dx_{2}, 
\end{equation}
where \(P(\boldsymbol{x}_{2}/\boldsymbol{x}_{1})\) is the conditional
probability density of finding any particle at $\boldsymbol{x}_{2}$
given
the presence of another particle at $\boldsymbol{x}_{1}$.

In appendix A we show, under several mild assumptions, that
\begin{equation}
\label{av100}
\left| \left< \boldsymbol{u} \right> - \bar{\boldsymbol{u}} - \left<
\boldsymbol{u}_{1} \right> \right| \sim O (c^{2}\log \frac{R}{\epsilon}
)
\end{equation}
and
\begin{equation}
\label{av110}
\left| \left< \boldsymbol{u} \right> - \bar{\boldsymbol{u}} - \left<
\boldsymbol{u}_{1} \right> - \left< \boldsymbol{u}_{2}\right> \right|
\sim
O (c^{3}\log^{2} \frac{R}{\epsilon})
\end{equation}
relative to $|\bar{\boldsymbol{u}}|$. The error estimates in
(\ref{av100})
and (\ref{av110}) arise respectively from the neglect of two- and
three-sphere hydrodynamic interactions. The presence of the term $\ln
R/\epsilon$
can be rationalized by the crudeness of the estimates derived in
Appendix
A. The fact that \(\left< \boldsymbol{u}_{2}\right>\) is of $O(c^{2})$
(as
will be shown later) supports this suggestion.

\newsec{Homogenization}
The goal of this section is to show that the average velocity
field obtained in the preceding section may be approximated by the
velocity field obtained for a homogeneous fluid
possessing a non-uniform viscosity, given the same boundaries and
ambient
flow field for the two cases. (The same can be shown for the comparable
stress fields, but the demonstration is omitted in the interests of
brevity.)

Our first step is to approximate the surface traction over the surface
of a
suspended sphere which is located far from the wall [i.e.,
$d(\boldsymbol{x}_{n},\partial \Omega) \gg \epsilon$, where
$\boldsymbol{x}_{n}$ denotes the location of the sphere's center]. Our
approximation represents the solution of the following problem:
\begin{equation}
\label{hom10}
\frac{1}{2} f_{0i}^{(n)}(\boldsymbol{y},\boldsymbol{x}_{n}) =
[2 \mu \bar{G}_{ij}(\boldsymbol{x}_{n}) -
\bar{p}(\boldsymbol{x}_{n})\delta_{ij}] \hat{n}_{j}(\boldsymbol{y}) +
\frac{3}{4 \pi} \int_{\partial s_{n}} \frac{r_{i} r_{j} r_{k}}{r^5}
f_{0k}^{(n)}(\boldsymbol{x,x}_{n})
ds_{x}\hat{n}_{j}(\boldsymbol{y}),
\end{equation}
wherein
\begin{displaymath}
G_{ij} = \frac{1}{2} \frac{\partial \bar{u}_{i}}{\partial x_{j}} +
\frac{\partial \bar{u}_{j}}{\partial x_{i}}. \end{displaymath}
Equation (\ref{hom10}) is almost identical with (\ref{av10}) except for
two
modifications: (a) neglect of the variation of $\boldsymbol{G}$
and $\bar{p}$ over $\partial s_{n}$, i.e., $\boldsymbol{G(y)} \cong
\boldsymbol{G(x}_{n})$ , \(\bar{p}(\boldsymbol{y}) \cong
\bar{p}(\boldsymbol{x}_{n})\);
(b) neglect of the Stokeslet image portion of the integral's kernel.
Item
(a) results in an $O(\epsilon /R)$ error, whereas (b) produces an
$(\epsilon^{3}/[d(\boldsymbol{x}_{n},\partial \Omega)]^{3})$ error,
which
means it is of $O(1)$ near the walls. Appendix B provides an estimate of
the overall effect of these neglections on the average velocity field.
Equation (\ref{hom10}) can readily be solved \cite{kika91}, to obtain
\begin{equation}
\label{hom20}
f_{0i}^{(n)} = 5 \mu G_{ij}(\boldsymbol{x}_{n})
\hat{n}_{j}(\boldsymbol{y}) -
\bar{p}(\boldsymbol{x}_{n})\hat{n}_{i}(\boldsymbol{y}). \end{equation}
Note that the portion of $\boldsymbol{f}_{0}$ due to shear has been
increased by a factor of $5/2$ relative to $\bar{\boldsymbol{f}}$. This
coefficient arises in the above context from the eigenvalue of the
integral
operator corresponding to shear flow over a sphere, which is $1/5$.

As is shown in appendix B, the error due to the items neglected above is
negligible even if we use (\ref{hom20}) everywhere (including the
vicinity
of the walls). Substitution into (\ref{av80}) yields
\begin{equation}
\label{hom30}
\left< u_{1i} \right> = - \frac{4}{3} \pi \epsilon^{3}
\int_{\Omega / B(\boldsymbol{x},\epsilon)} \left[ 1 + \frac{1}{10}
\epsilon^{2} \frac{\partial^2}{\partial x_{1p}^{2}} \right]
\frac{\partial}{\partial x_{1k}} \left[
T_{ij}(\boldsymbol{x}_{1},\boldsymbol{x}) \right] 5 \mu
G_{jk}(\boldsymbol{x}_{1})
n(\boldsymbol{x}_{1}) dx_{1} \; .
\end{equation}
The domain $B(\boldsymbol{x},\epsilon)$ has been omitted since
\begin{displaymath}
\int_{|\boldsymbol{\xi}|=\epsilon}
T_{ij}(\boldsymbol{x}_{1}+\boldsymbol{\xi},\boldsymbol{x})
f_{0j}^{(1)}(\boldsymbol{ \xi})ds_{\xi} \cong G_{ij} (x_{1j} - x_{j}) +
O(\frac{\epsilon}{R}) \; , \end{displaymath}
the $O(\epsilon/R)$ error being a consequence of our approximation
(\ref{hom20}) to $\boldsymbol{f}_{0}$. In view of (\ref{av80}) the
integral
over $B(\boldsymbol{x},\epsilon)$ vanishes. The $O(\epsilon^{2})$ term
appearing in (\ref{hom30}) also impacts negligibly upon $\left< u_{1i}
\right>$ since by applying the divergence theorem to it we obtain
\begin{gather}
\label{hom40}
\int_{\Omega / B(\boldsymbol{x},\epsilon)} \frac{1}{10} \epsilon^{2}
\frac{\partial^2}{\partial x_{1p}^{2}} \frac{\partial}{\partial x_{1k}}
\left[ T_{ij}(\boldsymbol{x}_{1},\boldsymbol{x}) \right] 5 \mu
G_{jk}(\boldsymbol{x}_{1})
n(\boldsymbol{x}_{1}) dx_{1} = \notag \\ \int_{\partial \Omega \bigcup
\partial
B(\boldsymbol{x},\epsilon)}\epsilon^{2}
\frac{\partial^2}{\partial x_{1p}^{2}}
\left[ T_{ij}(\boldsymbol{x}_{1},\boldsymbol{x}) \right] 5 \mu
G_{jk}(\boldsymbol{x}_{1})
n(\boldsymbol{x}_{1}) \hat{n}_{k} ds_{x_1} \; . \end{gather}The integral
over $\partial B(\boldsymbol{x},\epsilon)$ is
$O(\epsilon^3/R^3)$ since $\boldsymbol{T} \cong \boldsymbol{t}$ in that
domain. For those points $\boldsymbol{x}$ which lie well within the
interior of $\Omega$, i.e.,
$d(\boldsymbol{x},\partial \Omega) \sim O(R)$,we have that
$\boldsymbol{\nabla}
\boldsymbol{\nabla} \boldsymbol{T} \sim O(1/R^3)$. Consequently, the
above
integral is bounded by $C \epsilon^2/R^2 \| \boldsymbol{G} \| R
\sup_{\boldsymbol{x} \in
\Omega} n$. As such, it is then of $O(\epsilon^2/R^2)$ with respect to
the $O(1)$ term
in (\ref{hom30}).
Hence,
\begin{equation}
\label{hom60}
\left< u_{1i} \right> \cong - \frac{4}{3} \pi \epsilon^{3}
\int_{\Omega / B(\boldsymbol{x},\epsilon)} \frac{\partial}{\partial
x_{1k}}
\left[ T_{ij}(\boldsymbol{x}_{1},\boldsymbol{x}) \right] 5 \mu
G_{jk}(\boldsymbol{x}_{1})
n(\boldsymbol{x}_{1}) dx_{1},
\end{equation}
which upon applying the divergence theorem in conjunction with the fact
that $\boldsymbol{T}(\boldsymbol{x}_{1},\boldsymbol{x} )=0$ for
$\boldsymbol{x} \in \partial
\Omega_u$ yields (recall that $\hat{\boldsymbol n}$ is the {\em inward}
normal)
\begin{multline}
\label{hom70}
\left< u_{1i} \right> \cong \frac{4}{3} \pi \epsilon^{3}
\left\{ \int_{\Omega}
T_{ij}(\boldsymbol{x}_{1},\boldsymbol{x}) 5\mu \frac{\partial}{\partial
x_{1k}}\left[ G_{jk}(\boldsymbol{x}_{1}) n(\boldsymbol{x}_{1}) \right]
dx_{1} + \right. \\ \left. + \int_{\partial \Omega_f}
T_{ij}(\boldsymbol{x}_{1},\boldsymbol{x})5 \mu
G_{jk}(\boldsymbol{x}_{1})
n(\boldsymbol{x}_{1}) \hat{n}_{k} ds_{x_1} \right\} +
O(\frac{\epsilon}{R}).
\end{multline}
[We have ignored here the fact that
$n(\boldsymbol{x}_{1})=0$ for $d(\boldsymbol{x}_{1},\partial \Omega) <
\epsilon$, and instead assumed it to be smooth throughout the whole
domain.
It can easily be shown that the error produced by this assumption is of
$O(\epsilon/R)$.]
 
Our goal is to now obtain \( \left< \boldsymbol{u}_{1} \right> \) as a
solution of a Stokes problem in $\Omega$ for a homogeneous fluid
possessing
a non-uniform (but continuous) viscosity distribution, given the same
ambient flow in both cases. This Stokes problem has the following form:
\begin{subequation}
\begin{eqnarray}
\label{hom80}
& \displaystyle{\frac{\partial v_{i}}{\partial x_{i}} = 0} &
\mbox{for} \; x \in \Omega, \\
& \displaystyle{\frac{\partial }{\partial x_{j}} \left[ \mu_{s} \left(
\frac{\partial v_{i}}{\partial x_{j}} +
\frac{\partial v_{j}}{\partial x_{i}} \right) \right] = \frac{\partial
p}{\partial x_{i}}}
& \mbox{for} \; x \in \Omega, \\
& \displaystyle{\boldsymbol{v} = \bar{\boldsymbol{u}}} & \mbox{for} \; x
\in \partial \Omega_{u}, \\ &
\displaystyle{\boldsymbol{f}_{s}({\boldsymbol
v}) =
\boldsymbol{f}(\bar{\boldsymbol{u}})}
& \mbox{for} \; x \in \partial \Omega_f.
\end{eqnarray}
\end{subequation}
\hspace*{-1.5em}In the latter, ${\boldsymbol f}_s$ is determined by the
stress/rate-of-strain relation for the hypothetical homogeneous medium,
i.e.,
\begin{displaymath}
{\boldsymbol f}_s({\boldsymbol v}) = \left\{ \mu_{s} \left[ {\boldsymbol
\nabla}
{\boldsymbol v} + {\boldsymbol \nabla} {\boldsymbol v}^{\dagger} \right]
-
p{\boldsymbol I} \right\} \cdot \hat{\boldsymbol n}\; ,
\end{displaymath}
in contrast to ${\boldsymbol f}$, which is determined by the
stress/rate-of-strain relation for the Newtonian fluid (with $\mu_{s}$
replaced by $\mu$ in the above).
We seek a solution of (3.8) for circumstances in which $\mu_{s}$ is
nearly
constant. To this end we set
\begin{subequation}
\begin{eqnarray}
\label{hom90}
& & \mu_{s}(\boldsymbol{x}) = \mu \left\{ 1 + a_{1} \frac{4}{3} \pi
\epsilon^{3} n(\boldsymbol{x}) +a_{2} \left[ \frac{4}{3} \pi
\epsilon^{3}
n(\boldsymbol{x}) \right]^{2} + O(\epsilon^{9}) \right\}, \\
& & \boldsymbol{v} = \bar{\boldsymbol{u}} + \frac{4}{3} \pi \epsilon^{3}
\boldsymbol{v}_{1}
+ \left[ \frac{4}{3} \pi \epsilon^{3} \right]^{2} \boldsymbol{v}_{2}+
O(\epsilon^{9}), \\
& & p = \bar{p} + \frac{4}{3}
\pi \epsilon^{3} p_{1}
+ \left[ \frac{4}{3} \pi \epsilon^{3} \right]^{2} p_{2}+
O(\epsilon^{9}),
\end{eqnarray}
\end{subequation}
\hspace*{-1.5em}where $n(\boldsymbol{x})$ may be any positive continuous
function. The latter choice of notation derives from the fact that, as
will
subsquently be shown, in order to obtain equality between the velocity
field obtained from (3.8) and (3.9) and the respective average
velocity field obtained for the suspension, $n(\boldsymbol{x})$
must be identical
with the number density. Here, $\epsilon^{3}$ may be interpreted as the
order of the viscosity fluctuation, since we use the same dimensionless
spatial coordinate as before, and thus $n(\boldsymbol{x}) \sim O(1)$.
The
unknown constants $a_{1}$ and $a_{2}$ will be determined subsquently.
Upon substituting (3.9) into (3.8) the following boundary-value problem
is
obtained for $\boldsymbol{v}_{1}$: \begin{subequation}
\begin{eqnarray}
\label{hom100}
& & \frac{\partial v_{1i}}{\partial x_{i}} = 0, \\ & & \mu
\frac{\partial^{2} v_{1i}}{\partial x_{j}^{2}} - \frac{\partial
p_{1}}{\partial x_{i}} = -2a_{1} \mu \frac{\partial}{\partial x_{j}}
[nG_{ij}], \\ & & \left. \boldsymbol{v}_{1} \right|_{\boldsymbol{x}
\in \partial \Omega_u} = 0, \\
& & \left. \left[ \boldsymbol{f}_{1} + 2 a_{1} n({\boldsymbol x})
\mu {\boldsymbol G} \cdot \hat{\boldsymbol n} \right]
\right|_{\boldsymbol{x}
\in \partial \Omega_f} = 0 .
\end{eqnarray}
\end{subequation}
\hspace*{-1.5em} The solution of this set of equations may be expressed
in
the form of the following
integral representation \cite{habr83}:
\begin{equation}
\label{hom110}
v_{1i} = - \int_{\partial \Omega}
t_{ij}(\boldsymbol{x}_{1},\boldsymbol{x})
\left[ \mu \left( \frac{\partial v_{1j}}{\partial x_{1k}} +
\frac{\partial
v_{1k}}{\partial x_{1j}} \right) - p_{1} \delta_{jk} \right] \hat{n}_{k}
ds_{x_1} + 2a_{1} \mu \int_{\Omega}
t_{ij}(\boldsymbol{x}_{1},\boldsymbol{x}) \frac{\partial}{\partial
x_{k}}
[n G_{jk}] dx_{1}. \end{equation}
Using Green's theorem together with (3.10c) it is easy to show that if
we
replace the Stokeslet $\boldsymbol{t}$ appearing on the right-hand side
of
(\ref{hom110}) by its image $\boldsymbol{\tau}$, the left-hand side will
vanish. Hence,
\begin{equation}
\label{hom120}
v_{1i} = 2a_{1} \mu \int_{\Omega}
T_{ij}(\boldsymbol{x}_{1},\boldsymbol{x}) \frac{\partial}{\partial
x_{1k}}\left[ G_{jk}(\boldsymbol{x}_{1}) n(\boldsymbol{x}_{1}) \right]
dx_{1} + 2a_{1} \mu \int_{\partial \Omega_f}
T_{ij}(\boldsymbol{x}_{1},\boldsymbol{x}) G_{jk}(\boldsymbol{x}_{1})
n(\boldsymbol{x}_{1}) \hat{n}_{k} ds_{x_1}. \end{equation}
For the choice $a_{1}=5/2$, Eq.(\ref{hom70}) yields \begin{equation}
\label{hom130}
\left< \boldsymbol{u}_{1} \right> = \frac{4}{3} \pi
\epsilon^{3} \boldsymbol{v}_{1}.
\end{equation}
We have thus demonstrated that, to terms of $O(c)$, equality exists
between
the average
velocity field obtained for the suspension and the
velocity field obtained for a hypothetical homogeneous medium
characterized by the non-uniform viscosity field $\mu_{s} =
\mu [1+5/2 \: c(\boldsymbol{x})]$.

We next seek to obtain a comparable equality, but now up to terms of
$O(c^2)$, between the suspension-average and homogeneous fields. To this
end it is
necessary to separately discuss the respective contributions to the
average
velocity field of the relatively distant and relatively close sphere
pairs.
For the case of relatively distant pairs of spheres we may use the
approximation
\begin{eqnarray}
\label{hom150}
\left. \left( f_{0i}^{(n_{1},n_{2})} - f_{0i}^{(n_{1})} \right)
\right|_{\boldsymbol{x} \in \partial s_{n_1}} = - \frac{4}{3} \pi
\epsilon^{3} \left\{ \left( \frac{5}{2} \right)^{2} \mu^{2}
G_{jm}(\boldsymbol{x}_{n_2}) \cdot \nonumber \right. \\ \cdot
\frac{\partial}{\partial (x_{n_2})_{m}} \left. \left[ \frac{\partial
T_{ij}}{\partial (x_{n_1})_{k}} + \frac{\partial T_{kj}}{\partial
(x_{n_1})_{i}} \right]
\right|_{(\boldsymbol{x}_{n_2},\boldsymbol{x}_{n_1})}
\hat{n}_{k}(\boldsymbol{x}) - \nonumber \\ - \left. \left[ 5 \mu
G_{jk}(\boldsymbol{x}_{n_2}) - \bar{p}(\boldsymbol{x}_{n_2}) \delta_{jk}
\right] \left. \frac {\partial P_j}{\partial (x_{n_2})_{k}}
\right|_{(\boldsymbol{x}_{n_2},\boldsymbol{x}_{n_1})}\hat{n}_{i}(\boldsymbol{x})
\right\} \; , \end{eqnarray}
which is valid for \(|\boldsymbol{x}_{n_1} - \boldsymbol{x}_{n_2}| \gg
\epsilon \) . When \(|\boldsymbol{x}_{n_1} - \boldsymbol{x}_{n_2}| \sim
O(\epsilon) \), Eq.(\ref{av30}) has to be modified in the same manner
used
to obtain (\ref{hom10}), and the exact solution found. [We do not
however,
provide an error estimate, as previously done in appendix B in order to
justify (\ref{hom10})].
Substituting (\ref{hom150}) into (\ref{av90}) for
\(|\boldsymbol{x}_{n_1} -
\boldsymbol{x}_{n_2}| \geq L \gg \epsilon \) gives \begin{eqnarray}
\label{hom160}
\nopagebreak
& & \left< u_{2i} \right>(\boldsymbol{x}) \cong \left(
\frac{4}{3} \pi
\epsilon^{3} \right)^{2} \frac{5}{2}^{2} \mu^{2} \int_{\Omega /
B(\boldsymbol{x},\epsilon)} \frac{\partial}{\partial
x_{1k}} \left[ T_{ij}(\boldsymbol{x}_{1},\boldsymbol{x}) \right]
n(\boldsymbol{x}_{1})dx_{1} \cdot \nonumber \\ & & \cdot \int_{\Omega /
B(\boldsymbol{x}_{1},L)}
G_{pm}(\boldsymbol{x}_{2})\frac{\partial}{\partial
(x_{n_2})_{m}} \left. \left[
\frac{\partial T_{pj}}{\partial x_{1k}} + \frac{\partial
T_{kj}}{\partial
x_{1p}} \right]
\right|_{(\boldsymbol{x}_{2},\boldsymbol{x}_{1})}
n(\boldsymbol{x}_{2})dx_{2} +
\nonumber \\ & & + \int_{\Omega}n(\boldsymbol{x}_{1})dx_{1} \int_{2
\epsilon \leq \left| \boldsymbol{x}_{2} - \boldsymbol{x}_{1} \right|
\leq
L} \int_{|\boldsymbol{ \xi}|=\epsilon}
T_{ij}(\boldsymbol{x}_{1}+\boldsymbol{\xi},\boldsymbol{x})[
f_{0j}^{(1,2)}(\boldsymbol{x}_{1},\boldsymbol{x}_{2},\boldsymbol{\xi})-f_{0j}^{(
1)}(\boldsymbol{x}_{1},\boldsymbol{\xi})]ds_{\xi} \cdot \nonumber \\ & &
\cdot P(\boldsymbol{x}_{2}/\boldsymbol{x }_{1}) dx_{2} +
O(\frac{\epsilon^2}{L^2}),
\end{eqnarray}
where we have used two different properties of
\(P(\boldsymbol{x}_{2}/\boldsymbol{x}_{1})\), namely: $
P(\boldsymbol{x}_{2}/\boldsymbol{x}_{1}) \sim n$ for \linebreak
\(|\boldsymbol{x}_{1} - \boldsymbol{x}_{2}| \gg \epsilon \), and
$P(\boldsymbol{x}_{2}/\boldsymbol{x}_{1}) = 0$ for \(|\boldsymbol{x}_{1}
-
\boldsymbol{x}_{2}| < 2\epsilon \).

Upon applying the divergence theorem to the first term on the right-hand
side of (\ref{hom160}) and using the approximation 
\begin{equation}
\label{hom175}
\left. T_{ij}(\boldsymbol{x}_{1},\boldsymbol{x}_{2})
\right|_{|\boldsymbol{x}_{1} -
\boldsymbol{x}_{2}| = L} \cong
t_{ij}(\boldsymbol{x}_{1},\boldsymbol{x}_{2}) + O(\frac{L}{R}),
\end{equation}
we obtain
\begin{eqnarray}
\nopagebreak
\label{hom180}
& & \left< u_{2i} \right> (\boldsymbol{x}) \cong - \frac{4}{3} \pi
\epsilon^{3} 5 \mu \int_{\Omega} \frac {\partial
T_{ij}}{\partial x_{1k}} \left[ \frac{4}{3} \pi \epsilon^{3}
G_{jk}(\boldsymbol{x}_{1})n(\boldsymbol{x}_{1}) +
G_{1jk}(\boldsymbol{x}_{1}) \right]n(\boldsymbol{x}_{1}) dx_{1}
+ \nonumber \\
& & + \int_{\Omega}n^{2}(\boldsymbol{x}_{1})dx_{1} \int_{2 \epsilon \leq
\left| \boldsymbol{x}_{2} - \boldsymbol{x}_{1} \right| \leq L}
\int_{|\boldsymbol{ \xi}|=\epsilon}
T_{ij}(\boldsymbol{x}_{1}+\boldsymbol{\xi},\boldsymbol{x})[
f_{0j}^{(1,2)}(\boldsymbol{x}_{1},\boldsymbol{x}_{2},\boldsymbol{\xi})-f_{0j}^{(
1)}(\boldsymbol{x}_{1},\boldsymbol{\xi})]ds_{\xi} \cdot \nonumber \\ & &
\cdot q(\boldsymbol{x}_{2}/\boldsymbol{x }_{1}) dx_{2} +
O\left((\frac{\epsilon}{R}
)^{2/3}\right) \quad , \end{eqnarray}
wherein
\begin{equation}
\label{hom190}
G_{1jk} = \frac{1}{2}\left[
\frac{\partial \left< u_{1i} \right>}{\partial x_{j}} + \frac{\partial
\left< u_{1j} \right>}{\partial x_{i}} \right] \; ,
\end{equation}
$q(\boldsymbol{x}_{2}/\boldsymbol{x}_{1})=
P(\boldsymbol{x}_{2}/\boldsymbol{x}_{1})/n(\boldsymbol{x}_{1})$,  and
wherein $L$ was chosen to be $(\epsilon^{2}R)^{1/3}$ so as to minimize
the error.

To facilitate evaluation of the near-field term we discuss the domains
$|\boldsymbol{x}_{1} - \boldsymbol{x}| \geq \epsilon$ and
$|\boldsymbol{x}_{1} - \boldsymbol{x}| < \epsilon$ separately. For
$|\boldsymbol{x}_{1} - \boldsymbol{x}| \geq \epsilon$ we
expand $\boldsymbol{T}$ in power series of $\boldsymbol{\xi}$, i.e.,
\begin{equation}
\label{hom200}
T_{ij}(\boldsymbol{x}_{1}+\boldsymbol{\xi},\boldsymbol{x}) =
\sum_{n=0}^{\infty} \frac{1}{n!} \frac{\partial}{\partial (x_{1})_{m_1}}
\cdots \frac{\partial}{\partial (x_{1})_{m_n}}
T_{ij}(\boldsymbol{x}_{1},\boldsymbol{x}) \xi_{m_1} \cdots \xi_{m_n}.
\end{equation}
The next step entails obtaining $\boldsymbol{f}^{(1,2)}$ as a linear
combination of the $\boldsymbol{G}$ components: \begin{displaymath}
f_{0j}^{(1,2)} = A_{jlkm} G_{km} \hat{n}_{l}. \end{displaymath}
Observe that due to the special symmetry of the two-sphere geometry,
\begin{displaymath}
A_{jlkm}(\boldsymbol{x}_{2} -
\boldsymbol{x}_{1},\boldsymbol{\xi})
=A_{jlkm}^{\prime}(\boldsymbol{ R \cdot (x}_{2} -
\boldsymbol{x}_{1}),\boldsymbol{R\cdot \xi}), 
\end{displaymath}
where $\boldsymbol{R}$ is any rotation tensor, and the
$A_{jlkm}^{\prime}$
are the components of $\boldsymbol{A}$ in a coordinate system rotated by
$\boldsymbol{R}$; thus, 
\begin{equation}
\label{hom210}
A_{jlkm}(\boldsymbol{x}_{2} - \boldsymbol{x}_{1},\boldsymbol{\xi}) =
A_{jlkm}^{\prime}(\boldsymbol{x}_{1}
- \boldsymbol{x}_{2},-\boldsymbol{\xi})
=A_{jlkm}(\boldsymbol{x}_{1} -
\boldsymbol{x}_{2},\boldsymbol{\xi}).
\end{equation}
Consequently, the contributions of all the even terms in (\ref{hom200})
vanish if we assume
$q(\boldsymbol{x}_{2}/\boldsymbol{x}_{1}) =
q(2\boldsymbol{x}_{1}-\boldsymbol{x}_{2}/\boldsymbol{x}_{1})$, which
appears to be reasonable
hypothesis for $d(\boldsymbol{x}_{1},\partial \Omega) \gg \epsilon$;
(recall that $\left| \boldsymbol{x}_{2} - \boldsymbol{x}_{1} \right|
\leq L
$).

The contribution of the term corresponding to $n=1$ in (\ref{hom200}) is
\begin{equation}
\label{hom220}
- \int_{\Omega / B(\boldsymbol{x}, \epsilon)} n^{2}\frac {\partial
T_{ij}}{\partial x_{1k}}(\boldsymbol{x}_{1})dx_{1} \int_{2 \epsilon \leq
\left| \boldsymbol{x}_{2} - \boldsymbol{x}_{1} \right| \leq L} [
s_{jk}^{(1,2)}(\boldsymbol{x}_{1},\boldsymbol{x}_{2}) -
s_{jk}^{(1)}(\boldsymbol{x}_{1})]
q(\boldsymbol{x}_{2}/\boldsymbol{x}_{1})
dx_{2}, 
\end{equation}
wherein \( \boldsymbol{s}^{(1,2)} \) and \( \boldsymbol{s}^{(1)} \) are
the
stresslets of the sphere at $\boldsymbol{x}_{1}$ for the respective two-
and one-sphere problems.

The contributions of all other odd terms can be shown to be negligible:
applying the divergence theorem to the $2n+1$ term yields
\begin{gather}
\label{hom230}
\int_{\partial \Omega \bigcup \partial B(\boldsymbol{x},
\epsilon)} \frac{\partial}{\partial (x_{1})_{m_2}} \cdots
\frac{\partial}{\partial (x_{1})_{m_{2n+1}}}
T_{ij}(\boldsymbol{x}_{1},\boldsymbol{x}) \cdot \notag \\ \cdot
\int_{\Omega}
\int_{|\boldsymbol{\xi}| =\epsilon} \xi_{m_1} \cdots
\xi_{m_{2n+1}}[f_{0j}^{(1,2)}(\boldsymbol{x}_{1},\boldsymbol{x}_{2},
\boldsymbol{\xi})-f_{0j}^{(1)}(\boldsymbol{x}_{1},\boldsymbol{\xi})]
ds_{\xi}q(\boldsymbol{x}_{2}/\boldsymbol{x}_{1}) dx_{2}
n^{2}(\boldsymbol{x}_{1}) \hat{n}_{m_1} ds_{x_1} \quad . \end{gather}
Upon pursuing the same general arguments as those  following
(\ref{hom40}) we find that
the above term is of $O(\epsilon^{2n+1}/R^{2n+1})$.

Thus, upon combining Eqs.(\ref{hom180}) to (\ref{hom230}), and applying
the
divergence theorem once again, we obtain \begin{eqnarray}
\nopagebreak
\label{hom240}
& & \left< u_{2i} \right>(\boldsymbol{x}) \cong - \frac{4}{3} \pi
\epsilon^{3} 5 \mu \left[ \int_{\Omega} T_{ij} \frac
{\partial }{\partial x_{1k}} \left\{ \left[ \frac{4}{3} \pi
\epsilon^{3}G_{jk}(\boldsymbol{x}_{1})n(\boldsymbol{x}_{1}) +
G_{1jk}(\boldsymbol{x}_{1}) \right] n(\boldsymbol{x}_{1}) \right\}
dx_{1} +
\right. \nonumber \\ & & \left. + \int_{\partial \Omega_f} T_{ij}
\left( \frac{4}{3} \pi
\epsilon^{3}G_{jk}(\boldsymbol{x}_{1})n(\boldsymbol{x}_{1}) +
G_{1jk}(\boldsymbol{x}_{1}) \right) n(\boldsymbol{x}_{1}) \hat{n}_{k}
ds_{x_1} \right] + \nonumber \\ & & + \int_{\Omega} T_{ij} \frac
{\partial
}{\partial
x_{1k}} \left\{ \int_{2 \epsilon \leq \left| \boldsymbol{x}_{2} -
\boldsymbol{x}_{1} \right| \leq L} [
s_{jk}^{(1,2)}(\boldsymbol{x}_{1},\boldsymbol{x}_{2}) -
s_{jk}^{(1)}(\boldsymbol{x}_{1})]
q(\boldsymbol{x}_{2}/\boldsymbol{x}_{1})
dx_{2}n^{2}(\boldsymbol{x}_{1}) \right\} dx_1 + \nonumber \\ & & +
\int_{\partial \Omega_f} T_{ij} \int_{2 \epsilon \leq \left|
\boldsymbol{x}_{2} - \boldsymbol{x}_{1} \right| \leq L} [
s_{jk}^{(1,2)}(\boldsymbol{x}_{1},\boldsymbol{x}_{2}) -
s_{jk}^{(1)}(\boldsymbol{x}_{1})]
q(\boldsymbol{x}_{2}/\boldsymbol{x}_{1})
dx_{2}n^{2}(\boldsymbol{x}_{1}) \hat{n}_{k} ds_{x_1} + \nonumber \\ & &
+
n^{2}(\boldsymbol{x})
\int_{|\boldsymbol{x}_{1} - \boldsymbol{x}| <
\epsilon} \int_{2 \epsilon
\leq \left| \boldsymbol{x}_{2} - \boldsymbol{x}_{1} \right| \leq L}
\int_{|\boldsymbol{ \xi}|=\epsilon}
T_{ij}(\boldsymbol{x}_{1}+\boldsymbol{\xi},\boldsymbol{x})[
f_{0j}^{(1,2)}(\boldsymbol{x}_{1},\boldsymbol{x}_{2},\boldsymbol{\xi})-f_{0j}^{(
1)} (\boldsymbol{x}_{1},\boldsymbol{\xi})]ds_{\xi} \cdot \nonumber \\ &
&
\cdot q(\boldsymbol{x}_{2}/\boldsymbol{x}_{1}) dx_{2} dx_{1} +
O(\frac{\epsilon}{L}^{2}). \end{eqnarray}
Utilizing (\ref{hom210}) it is easy to show that the last term on the
right-hand side of (\ref{hom240}) vanishes by applying the following
transformation:
\begin{eqnarray*}
\boldsymbol{\xi} \to -\boldsymbol{\xi} \hspace{0.6cm} ; &
\boldsymbol{x}_{1} \to 2\boldsymbol{x} - \boldsymbol{x}_{1} & ;
\hspace{0.6cm} \boldsymbol{x}_{2}
\to 2\boldsymbol{x}_{1} - \boldsymbol{x}_{2} \; . \; \;
\mbox{Q.E.D} \; .
\end{eqnarray*}

The function $q(\boldsymbol{x}_{2}/\boldsymbol{x}_{1})$ obtained by
Batchelor \& Green \cite{bagr72b} for elongational flows considers only
the
effect of convection; [for simple shear flow the effect of convection
alone
is insufficient to determine
$q(\boldsymbol{x}_{2}/\boldsymbol{x}_{1})$].
For the case of elongational flows, \begin{eqnarray}
\label{hom250}
& & \int_{2 \epsilon \leq \left| \boldsymbol{x}_{2} -
\boldsymbol{x}_{1} \right|
\leq L} [ s_{jk}^{(1,2)}(\boldsymbol{x}_{1},\boldsymbol{x}_{2}) -
s_{jk}^{(1)}(\boldsymbol{x}_{1})]
q(\boldsymbol{x}_{2}/\boldsymbol{x}_{1})
dx_{2} \cong \nonumber \\ & & \cong \left( \frac{4}{3} \pi
\epsilon^{3} \right)^{2} 2 \mu G_{jk}(\boldsymbol{x}_{1}) \frac{15}{2}
\int_{2}^{\infty} J(\zeta)q(\zeta)d\zeta + O(\frac{L}{R}) \cong
\nonumber
\\ & & \cong 4.45 \left( \frac{4}{3} \pi \epsilon^{3} \right)^{2} 2 \mu
G_{jk}(\boldsymbol{x}_{1}), \end{eqnarray}
where $J$ is a hydrodynamic coefficient defined in \cite{bagr72a} and
\cite{bagr72b}. The coefficient 4.45 is taken from the numerical
calculations of Yoon \& Kim \cite{yoki87}. Substituting in
(\ref{hom240})
we obtain \begin{eqnarray}
\label{hom260}
& & \left< u_{2i} \right>(\boldsymbol{x}) \cong \frac{4}{3} \pi
\epsilon^{3} \left\{ 5 \mu \left[ \int_{\Omega} T_{ij} \frac{\partial
}{\partial x_{1k}} \left[ G_{1jk}(\boldsymbol{x}_{1})
n(\boldsymbol{x}_{1})\right] dx_{1} + \right. \right. \nonumber \\ & &
\left. + \int_{\partial \Omega_f} T_{ij}
G_{1jk}(\boldsymbol{x}_{1}) n(\boldsymbol{x}_{1}) \hat{n}_k ds_{x_1}
\right] + \nonumber \\ & & + \; (\frac{5}{2} + 4.45) 2 \mu \left[
\int_{\Omega} T_{ij} \frac
{\partial }{\partial x_{1k}} \left( \frac{4}{3} \pi
\epsilon^{3}G_{jk}(\boldsymbol{x}_{1})
n^{2}(\boldsymbol{x}_{1})\right) dx_{1} \right. \nonumber \\ & & \left.
\left. + \int_{\partial \Omega_f} T_{ij} \frac{4}{3} \pi
\epsilon^{3} G_{jk}(\boldsymbol{x}_{1})
n^{2}(\boldsymbol{x}_{1}) \hat{n}_{k} ds_{x_1}\right] \right\} .
\end{eqnarray}
The $5/2$ coefficient appearing in the second integral on the right-hand
side
arises from far-field, two-sphere interactions. Batchelor \& Green
\cite{bagr72b} arrive at the same conclusion by considering the somewhat
artificial situation $q(\boldsymbol{x}_{2}/\boldsymbol{x}_{1})=0$ for
$\left|\boldsymbol{x}_{1} - \boldsymbol{x}_{2} \right| \leq L$. Here, it
is
derived more naturally through the separation of far-field and
near-field
contributions.

To obtain $O(c^2)$ equality between the suspension-average velocity
field
and the velocity field obtained for a homogeneous fluid possessing a
non-uniform viscosity, we consider the next higher-order balance
equations for the homogeneous continuum problem, namely
\begin{subequation}
\begin{eqnarray}
\label{hom270}
& & \frac{\partial v_{2i}}{\partial x_{i}} = 0, \\ & & \mu
\frac{\partial^{2} v_{2i}}{\partial x_{j}^{2}} - \frac{\partial
p_{2}}{\partial x_{i}} = -5 \mu \frac{\partial}{\partial x_{j}}
[nG_{1ij}^{\prime}] - 2a_{2} \mu \frac{\partial}{\partial x_{j}}
[n^{2}G_{ij}], \\ & & \left. \boldsymbol{v}_{1} \right|_{\boldsymbol{x}
\in \partial \Omega_u} = 0, \\
&& \left. \left\{ \boldsymbol{f}_{2} + \left[ 5 n({\boldsymbol x})
\mu {\boldsymbol G}_{1}^{\prime} + 2 a_{2} n({\boldsymbol x})\mu
{\boldsymbol G} \right] \cdot \hat{\boldsymbol n} \right\}
\right|_{\boldsymbol{x} \in \partial \Omega_f} = 0 ,
\end{eqnarray}
\end{subequation}
\hspace*{-1.5em} wherein $(4/3)\: \pi \epsilon^3 {\boldsymbol
G}_{1}^{\prime} = {\boldsymbol G}_{1}$ Similarly to (\ref{hom120}) we
obtain
\begin{eqnarray}
\label{hom280}
& & v_{2i} (\boldsymbol{x}) = 5 \mu \left\{
\int_{\Omega} T_{ij} \frac {\partial }{\partial x_{1k}} \left[
G_{1jk}^{\prime}(\boldsymbol{x}_{1}) n(\boldsymbol{x}_{1})\right] dx_{1}
+
\int_{\partial \Omega_f} T_{ij} G_{1jk}^{\prime}(\boldsymbol{x}_{1})
n(\boldsymbol{x}_{1}) \hat{n}_k ds_{x_1} \right\} + \nonumber \\ & & + 2
a_{2} \mu \left\{ \int_{\Omega} T_{ij} \frac
{\partial }{\partial x_{1k}} \left[ G_{jk}(\boldsymbol{x}_{1})
n^{2}(\boldsymbol{x}_{1})\right] dx_{1} + \int_{\partial
\Omega_f} T_{ij} G_{jk}(\boldsymbol{x}_{1}) n^{2}(\boldsymbol{x}_{1})
\hat{n}_{k} ds_{x_1} \right\} .
\end{eqnarray}
Thus, the choice $a_{2}=6.95$ yields
\begin{equation}
\label{hom290}
\left< \boldsymbol{u}_{2} \right> = \left( \frac{4}{3} \pi
\epsilon^{3} \right)^{2} \boldsymbol{v}_{1}, \end{equation}
thereby establishing an $O(c^2)$ equality between the respective
velocity
fields.

\newsec{Concluding remarks}
This section addresses several key points insufficiently emphasized in
the
prior analysis:

1. {\em The error:} Two types of errors are produced in the
process of approximation. The first arises from neglecting higher-order
hydrodynamic interactions (e.g., three-sphere interactions in the case
when
the expansion addresses only two-sphere interactions) and is of
$O(c^{n+1})$, where $n$ refers to the $n$-sphere interactions explicitly
considered. The second error stems from ignoring 'global' effects
(arising
from the presence of boundaries
and spatial inhomogeneities) on the surface traction, and is of $O\left(
c\epsilon/R,c^{2}(\epsilon/R)^{(2/3} \right)$. Thus,
there is no point in considering
more than $n$ terms if $c^n \sim O\left(
(\epsilon/R),c(\epsilon/R)^{(2/3)}
\right)$. We may conclude, therefore, that for a dilute suspension the
required number of terms in our expansion is rather small, whereas for a
more concentrated suspension it might be useful to obtain more terms
(assuming, of course, that the expansion is convergent, an issue which
is
not yet clear).

2. {\em Comparison with the local approach:} The major difference
between
the {\em global} approach, presented in the present work, and the
classical
{\em local} one (cf. \cite{bagr72a}, \cite{bagr72b}, \cite{hi77},
\cite{ob79}) resides in
the order in which limits are being taken. Local analysis assumes {\em a
priori} that global effects are unimportant, i.e., that the medium is
unbounded in extent and that the number density and ambient
rate-of-strain
are spatially uniform. Mathematically, this means that the ratio of
particle size to macroscopic
length scale, $\epsilon/R$, is set to be equal to zero {\em before}
performing any further analysis. This limit process, which is certainly
an
idealization, was based in prior analyses on the {\em hope} that in the
real physical systems encountered in practice, neither the presence of
boundaries nor of spatial inhomogeneties would have any effect on the
suspension viscosity, which was intuitively expected to be a local
quantity. In contrast to the local approach, we first calculate the
velocity field -- including the global effects -- and only then do we
pass
to the above limit. Thereby, we show explicitly that 
\begin{displaymath}
\left| \left< {\boldsymbol u} \right> - \bar{\boldsymbol u} -
\frac{4}{3}\epsilon^3 {\boldsymbol v}_{1} - (\frac{4}{3}\epsilon^3)^2
{\boldsymbol v}_{2} \right| \leq C \sup_{{\boldsymbol x} \in \Omega}
\left|
\bar{\boldsymbol u} \right| \left[ c \frac{\epsilon}{R} c^{2} \left(
\frac{\epsilon}{R} \right)^{2/3} + c^3 \right] , \end{displaymath}
thus demonstrating that the suspension is equipollent, on average, to a
homogeneous medium, possessing a non-uniform viscosity field. It should
be
emphasized that {\em this result is independent of the domain's size and
shape}.

In addition for providing an estimate for the error generated by the
presence of boundaries and spatial inhomogeneities, global analysis
possesses another significant advantage over the local approach: It
avoids
the classical Stokesian divergence problem, and hence the need for
renormalization \cite{bagr72a}, \cite{bagr72b}, \cite{hi77},
\cite{ob79}!
It is well known that
the Stokesian divergence issue poses serious problems when seeking
physically meaningful solutions in unbounded domains, for both
suspensions
and homogeneous fluids. In fact, had we worked in $R^3$, many of the
integrals in sections 2 and 3 would have been divergent. The best way to
avoid the Stokesian divergence is to discuss large but finite domains,
which is exactly what was done in Ref. \cite{ruke89a} for the
sedimentation
case, and what has been done here in the present paper for the
suspension
rheology case.

\vspace{1ex}
\begin{center}
{\bf Acknowledgements}
\end{center}
\vspace{1ex}
\nopagebreak[4]
Y.A. wishes to thank the Fulbright Scholar Program for their support
during
the tenure of this research. H.B. was supported by a grant from the
Office
of Basic Energy Sciences of the U.S. Department of Energy.

\appsec{Error estimates for one- and two-sphere modes} 
The goal of this
appendix is to demonstrate the validity of (\ref{av100}) and
(\ref{av110}).
To this end, it is first necessary to prove uniform boundedness of
$\boldsymbol{f}$. 
\begin{lemma}
\begin{displaymath}
\exists \epsilon_{0}>0 \; \text{such that} \; \|\boldsymbol{f}
\|_{\displaystyle L^{\infty} \scriptstyle
\left[\bigcup_{n=1}^{N} \partial s_{n}\right]} \leq C(N,\Omega) \mu \|
\boldsymbol{G} \| \; \forall  0<\epsilon \leq \epsilon_{0} \; \text{and}
\;
(\boldsymbol{x}_{1},\ldots,\boldsymbol{x}_{N}) \in D^{N}
\end{displaymath}
for every $D \subset \Omega$
satisfying $d(\partial D,\partial \Omega) \geq \epsilon$ and for smooth
$\partial \Omega$.
\end{lemma}

The proof of lemma 1 \cite{lenote} utilizes the fact that
$\boldsymbol{f}$ tends to the solution of $N$-particle problem in
$\Bbb{R}^{3}$. 
Yet, it is still necessary to show that $\boldsymbol{f}$ remains uniformly
bounded even if some of the particles are allowed to approach the
boundary proximities. To this end we first {\em assume} that
$\boldsymbol{f}$ tends, as 
$\epsilon \to 0$ and for $d(\boldsymbol{x}_{n},\partial
\Omega) \leq C \epsilon \quad \forall 1 \leq n \leq N$ to the
solution of $N$-particle problem near a flat wall (which looks
intuitively correct, but seems to be technically difficult to
prove). Then, similar arguments to those presented in the proof
of lemma 1 (where \cite{lenote} can be applied in order to
demonstrate uniform boundedness in this case as well. 

The above uniform boundedness property suffices, as the
subsequent analysis demonstrates, to prove 
(\ref{av100}) and (\ref{av110}) for fixed $N$ as $\epsilon \to
0$.
Yet, a much more interesting limiting case arises when both $N
\to \infty$
and $\epsilon \to 0$, for fixed $N/R^3$. In terms of the
original physical
variables, this case arises if we either let the boundaries approach
infinity, or else let the average distance between particles tend to
zero. Uniform boundedness of
$\boldsymbol{f}$ in that limit seems, however, to be difficult to prove.
It
would perhaps be easier to prove that the average of $\boldsymbol{f}$
with
respect to $N-3$ particle locations is bounded, i.e.,
\begin{equation}
\label{A05}
\int \boldsymbol{f}
f_{N}(\boldsymbol{x}_{1},\ldots,\boldsymbol{x}_{N}) dx_{4} \cdots dx_{N}
\leq C(\Omega) \mu \| \boldsymbol{G} \|
f_{3}(\boldsymbol{x}_{1},\boldsymbol{x}_{2},\boldsymbol{x}_{3}),
\end{equation}
which, if correct, suffices in order to prove both (\ref{av100}) and
(\ref{av110}. We shall, thus {\em assume} the validity of (\ref{A05}) in
the limit $N
\to \infty$. Further research is, however, necessary in order to
prove this assumption.

Before deriving (\ref{av100}) and (\ref{av110}) we prove the following
auxiliary result:
\begin{lemma}
\begin{equation}
\label{A06}
\sup_{\boldsymbol{y} \in \partial s_{n_1}} \left| \int_{\partial
s_{n_2}} \left. \boldsymbol{\sigma}_{y}(\boldsymbol{T})
\right|_{(\boldsymbol{x},\boldsymbol{y})} \cdot
\boldsymbol{f}(\boldsymbol{x}) ds_{x} \cdot
\hat{\boldsymbol{n}}(\boldsymbol{y}) \right| \leq
C \epsilon^{3} \frac{\|\boldsymbol{f}\|_{L^{\infty}(\partial
s_{n_2})}} {|\boldsymbol{x}_{n_1} -
\boldsymbol{x}_{n_2}|^{3}} \; .
\end{equation}
\end{lemma}
{\em proof:} Use of the fact that all spheres are neutrally buoyant
furnishes the inequality \begin{asubequation}
\label{A4}
\begin{eqnarray}
& \displaystyle {\sup_{\boldsymbol{y} \in \partial s_{n_1}}
\left| \int_{\partial
s_{n_2}} \left. \boldsymbol{\sigma}_{y}(\boldsymbol{T})
\right|_{(\boldsymbol{x},\boldsymbol{y})} \cdot
\boldsymbol{f}(\boldsymbol{x}) ds_{x} \cdot
\hat{\boldsymbol{n}}(\boldsymbol{y})
\right| \leq} & \nonumber \\ & \displaystyle \leq \left\|
(\boldsymbol{x-x}_{n}) \cdot
\nabla_{x}\left. \boldsymbol{\sigma}_{y}(\boldsymbol{T})
\right|_{(\boldsymbol{x}^{\prime},\boldsymbol{y})} \cdot
\hat{\boldsymbol{n}}(\boldsymbol{y}) \right\|_{L^{1}(\partial s_{n_2})}
\|
\boldsymbol{f} \|_{L^{\infty}(\partial
s_{n_2})} & ,
\end{eqnarray}
wherein
\begin{eqnarray}
& \boldsymbol{x}^{\prime} =
\boldsymbol{x}_{n} + \theta(\boldsymbol{ x-x}_{n}), \; \; \; \;
(0 \leq \theta \leq 1).
\end{eqnarray}
\end{asubequation}
\hspace*{-1.5em} It can be shown \cite{la63} that \begin{equation}
\label{A5}
\left| \nabla_{x} \left.
\boldsymbol{\sigma}_{y}(\boldsymbol{T})\right|_{(\boldsymbol{x}^{\prime},
\boldsymbol{y})} \right| \leq
\frac{C}{|\boldsymbol{x}^{\prime}-\boldsymbol{y}|^{3}}, \hspace{1cm}
(\boldsymbol{x}^{\prime},\boldsymbol{y}) \in \Omega. \end{equation}
Consequently, (\ref{A06})holds for
$|\boldsymbol{x}_{n_2}-\boldsymbol{x}_{n_1}| \geq 3 \epsilon$. For
$|\boldsymbol{x}_{n_2}-\boldsymbol{x}_{n_1}| < 3 \epsilon$ it is easy to
show that for
$\boldsymbol{x} \neq \boldsymbol{y} $,
\begin{equation}
\label{A7}
\left| (\boldsymbol{x-x}_{n}) \cdot \nabla_{x}\left.
\boldsymbol{\sigma}_{y}(\boldsymbol{T})
\right|_{(\boldsymbol{x}^{\prime},\boldsymbol{y})} \cdot
\hat{\boldsymbol{n}}(\boldsymbol{y}) \right| \leq C_{1} \left|
(\boldsymbol{x-x}_{n}) \cdot\nabla \left. \frac{\boldsymbol{ rrr}}{r^5}
\right|_{(\boldsymbol{r=x}-\boldsymbol{y})} \cdot
\hat{\boldsymbol{n}}(\boldsymbol{y}) \right| + \frac{C_{2}}{r} \; .
\end{equation}
Hence, since
\begin{displaymath}
\stackrel{\displaystyle sup}{\scriptstyle y \in \partial s_{n_2}}
\left\|
(\boldsymbol{x-x}_{n}) \cdot\nabla_{x}
\left. \frac{\boldsymbol{ rrr}}{r^5}
\right|_{(\boldsymbol{r=x}-\boldsymbol{y})} \cdot
\hat{\boldsymbol{n}}(\boldsymbol{y}) \right\|_{L^{1}(\partial s_{n_2})}
\end{displaymath}
is finite, (\ref{A06}) remains valid for $2\epsilon \leq
|\boldsymbol{x}_{n_2}-\boldsymbol{x}_{n_1}| < 3 \epsilon$ as well.
\begin{displaymath}
\square
\end{displaymath}
In the reminder of this appendix we derive first the one-sphere
approximation (\ref{av100}), followed subsequently by the two-sphere
approximation
(\ref{av110}).
\pagebreak[0]

{\em One-sphere case:} The equation for
$\boldsymbol{f-f}_{0}^{(n)}$ can be obtained via (\ref{int30}) and
(\ref{av10}) as
\begin{eqnarray}
\label{A1}
\frac{1}{2} (f_{i}-f_{0i}^{(n_{1})})
(\boldsymbol{y},\boldsymbol{x,x}_{1},\ldots,\boldsymbol{x}_{N}) & = &
\bar{f}_{i}(\boldsymbol{y}) +
\int_{\partial s_{n_1}} (\sigma_{y})_{ij}(\boldsymbol{T}_{.k}) (f_{k} -
f_{0k}^{(n_{1})})(\boldsymbol{x,x}_{1},\ldots,\boldsymbol{x}_{N})
ds_{x}\hat{n}_{j}(\boldsymbol{y}) + \nonumber \\ & & + \sum_{\stackrel
{\scriptstyle n_{2} = 1} {n_{2} \neq n_{1}}} ^{N} \int_{\partial
s_{n_2}}
(\sigma_{y})_{ij}(\boldsymbol{T}_{.k})
f_{k}(\boldsymbol{x,x}_{n_{1}}, \ldots \boldsymbol{x}_{N}) ds_{x}
\hat{n}_{j}(\boldsymbol{y}).
\end{eqnarray}

Inasmuch as (\ref{A1}) is a Fredholm equation, use of (\ref{A06}) gives
\begin{equation}
\label{A8}
\| \boldsymbol{f-f}_{0}^{n_1} \|_{L^{\infty}(\partial s_{n_1})} \leq C
\sum_{\stackrel {\scriptstyle n_{2} = 1} {n_{2} \neq
n_{1}}}^{N} \frac{\| \boldsymbol{f} \|
\epsilon^3}{|\boldsymbol{x}_{n_2}-\boldsymbol{x}_{n_1}|^{3}}.
\end{equation}
The error in the configuration-dependent velocity is \begin{equation}
\label{A9}
\Delta u_{1i} = \sum_{n=1}^{N} \int_{\partial s_{n}}
T_{ij}(f_{j}-f_{0j}^{n}) ds_{x} \; .
\end{equation}
Invoking the fact
that \cite{la63}
\begin{displaymath}
\left| \nabla_{x} \left. \boldsymbol{T}
\right|_{(\boldsymbol{x},\boldsymbol{y})} \right| \leq
\frac{C}{|\boldsymbol{x}-\boldsymbol{y}|^{2}} \; , \end{displaymath}
we obtain
\begin{equation}
\label{A10}
| \Delta \boldsymbol{u}_1 | \leq \frac{C}{\mu} \sum_{n=1}^{N}
\frac{\|\boldsymbol{f}\|
\epsilon^{6}}{|\boldsymbol{x}_{n_2}-\boldsymbol{x}_{n_1}|^{3}
|\boldsymbol{x}_{n_1}-\boldsymbol{x}|^{2}} \; . \end{equation}
Assuming that (\ref{A05}) is valid, the error in the average velocity is
bounded by
\begin{equation}
\label{A11}
\left|< \boldsymbol{u} -\bar{\boldsymbol{u}} -
\boldsymbol{u}_{1} \right>| \leq
\epsilon^{6} C
\| \boldsymbol{G} \| \int_{\Omega}
\frac{n(\boldsymbol{x}_{1})
dx_{1}}{|\boldsymbol{x}_{1}-\boldsymbol{x}|^{2}} \int_{\Omega}
\frac{P(\boldsymbol{x}_{2}/\boldsymbol{x}_{1}) dx_{2}}
{|\boldsymbol{x}_{2}-\boldsymbol{x}_{1}|^{3}}, \end{equation}
where $C$ is independent of $N$. (For fixed $N$ the above estimate
follows
from lemma 1.) Since \( P(\boldsymbol{x}_{2}/\boldsymbol{x}_{1}) \sim
n(\boldsymbol{x}_{1}) \) for
$|\boldsymbol{x}_{2}-\boldsymbol{x}_{1}| \gg \epsilon$, it can obviously
be
asserted that \begin{displaymath}
P(\boldsymbol{x}_{2}/\boldsymbol{x}_{1}) \leq C_{1}n(\boldsymbol{x}_{1})
\hspace{1cm} \forall |\boldsymbol{x}_{2}-\boldsymbol{x}_{1}| \geq
C_{2}\epsilon,
\end{displaymath}
where $C_{1}$ and \( C_{2} \) are constants of $O(1)$, i.e., independent
of
$\epsilon$. Obviously, since \begin{displaymath}
\int_{|\boldsymbol{x}_{2}-\boldsymbol{x}_{1}| \leq C
\epsilon } P(\boldsymbol{x}_{2}/\boldsymbol{x}_{1}) dx_{2} \cong
\int_{|\boldsymbol{x}_{2}-\boldsymbol{x}_{1}| \leq C \epsilon }
n(\boldsymbol{x}_{2}) dx_{2}
\end{displaymath}
for $C \gg 1$, we may assert that
\begin{displaymath}
\int_{|\boldsymbol{x}_{2}-\boldsymbol{x}_{1}| \leq
C_{2} \epsilon }
\frac{P(\boldsymbol{x}_{2}/\boldsymbol{x}_{1}) dx_{2}}
{|\boldsymbol{x}_{2}-\boldsymbol{x}_{1}|^{3}} \leq \frac{1}{2
\epsilon^3}
\int_{|\boldsymbol{x}_{2}-\boldsymbol{x}_{1}| \leq
C_{2} \epsilon } P(\boldsymbol{x}_{2}/\boldsymbol{x}_{1}) dx_{2} \leq
C_{3}
n(\boldsymbol{x}_{1}).
\end{displaymath}

In the above we have implicitly assumed that \( P(\alpha
(\boldsymbol{x}_{2}-\boldsymbol{x}_{1})/\boldsymbol{x}_{1}, \alpha
\epsilon) \leq \linebreak
CP(\boldsymbol{x}_{2}-\boldsymbol{x}_{1}/\boldsymbol{x}_{1},\epsilon)
\) for all $0 < \alpha \leq 1$, where $C$ is independent of $\alpha$,
$\epsilon$, and $N$. Such an assumption is clearly necessary in order to
guarantee the boundedness of $P(\boldsymbol{x}_{2}/\boldsymbol{x}_{1})$
in the
limit $N \to \infty$, $\epsilon \to 0$. The
approximations
we derive will thus be valid only in cases where the pair probability
behaves according to the above assumption. It is expected, however, that
in
most cases $P(\boldsymbol{x}_{2}/\boldsymbol{x}_{1})$ will satisfy this
assumption, since it obeys simple conservation laws which do not
depend upon the $\epsilon$ and $N$ values. (cf. Ref.\cite{bagr72b} for
instance).

Consequently,
\begin{equation}
\label{A12}
\int_{\Omega} \frac{P(\boldsymbol{x}_{2}/\boldsymbol{x}_{1}) dx_{2}}
{|\boldsymbol{x}_{2}-\boldsymbol{x}_{1}|^{3}} \leq
\bar{C}\int_{|\boldsymbol{x}_{2} - \boldsymbol{x}_{1}| \geq 2 \epsilon}
\frac{n(\boldsymbol{x}_{2}) dx_{2}}
{|\boldsymbol{x}_{2}-\boldsymbol{x}_{1}|^{3}}. \end{equation}
Hence, since $n(\boldsymbol{x}_{2}) \leq n_{max} \sim O(1)$, one obtains
via
(\ref{A11}) and (\ref{A12}) the inequality \begin{equation}
\label{A13}
| \left< \boldsymbol{u} -\bar{\boldsymbol{u}} -
\boldsymbol{u}_{1} \right> | \leq
\epsilon^{6} n_{max}^{2}C
\| \boldsymbol{G} \| R \ln \left(
\frac{R}{\epsilon} \right) ,
\end{equation}
a result which coincides with (\ref{av100}).

\vspace{1ex}
{\em Two-sphere case:} Similarly to (\ref{A1}), one may obtain the
equation
\begin{eqnarray}
\label{A14}
& & \frac{1}{2} (f_{i}
-f_{0i}^{(n_{1},n_{2})})(\boldsymbol{y},\boldsymbol{x,x}_{1},\ldots,
\boldsymbol{x}_{N})
= \bar{f}_{i}(\boldsymbol{y}) + \nonumber \\ & & + \int_{\partial
s_{n_1}
\bigcup \partial s_{n_2}}
(\sigma_{y})_{ij}(\boldsymbol{T}_{.k}) (f_{k} -
f_{0k}^{(n_{1},n_{2})})(\boldsymbol{x,x}_{1},\ldots,\boldsymbol{x}_{N})
ds_{x}\hat{n}_{j}(\boldsymbol{y}) + \nonumber \\ & & + \sum_{\stackrel
{\scriptstyle n_{3} = 1} {n_{3} \neq n_{1},n_{2}}}
^{N} \int_{\partial s_{n}}
(\sigma_{y})_{ij}(\boldsymbol{T}_{.k})
f_{k}(\boldsymbol{x,x}_{n_{1}}, \ldots \boldsymbol{x}_{N}) ds_{x}
\hat{n}_{j}(\boldsymbol{y}).
\end{eqnarray}
Application of the same procedure used to obtain (\ref{A8}) gives, with
the
aid of lemma 1,
\begin{equation}
\label{A15}
\| \boldsymbol{f-f}_{0}^{(n_{1},n_{2})} \|_{L^{\infty}(\partial s_{n_1}
\bigcup \partial s_{n_2})} \leq C
\mu \| \boldsymbol{G} \| \epsilon^3 \sum_{\stackrel
{\scriptstyle n_{3}
= 1} {n_{3} \neq n_{1},n_{2}}}^{N} \left[
\frac{1}{|\boldsymbol{x}_{n_3}-\boldsymbol{x}_{n_2}|^{3}} +
\frac{1}{|\boldsymbol{x}_{n_3}-\boldsymbol{x}_{n_1}|^{3}} \right].
\end{equation}
We seek an estimate for $\| \boldsymbol{f-f}_{0}^{(n_{1},n_{2})}
\|_{L^{\infty}(\partial s_{n_2})}$. We therefore first interpret
(\ref{A14}) as a Fredholm equation over $\partial s_{n_2}$ to obtain
\begin{eqnarray}
\samepage
\label{A16}
\| \boldsymbol{f-f}_{0}^{(n_{1},n_{2})} \|_{L^{\infty}(\partial
s_{n_2})}
& \leq & C \mu \| \boldsymbol{G} \| \epsilon^3 \sum_{\stackrel
{\scriptstyle n_{3} = 1} {n_{3} \neq n_{1},n_{2}}}^{N}
\frac{1}{|\boldsymbol{x}_{n_3}-\boldsymbol{x}_{n_2}|^{3}} + \nonumber \\
&
& + \frac{\epsilon^3}{|\boldsymbol{x}_{n_2}-\boldsymbol{x}_{n_1}|^{3}}\|
\boldsymbol{f-f}_{0}^{(n_{1},n_{2})}
\|_{L^{\infty}(\partial s_{n_1})}.
\end{eqnarray}
In combination, Eqs.(\ref{A15}) and (\ref{A16}) give \begin{equation}
\label{A17}
\| \boldsymbol{f-f}_{0}^{(n_{1},n_{2})} \|_{L^{\infty}(\partial
s_{n_2})} \leq C \mu \| \boldsymbol{G} \| \epsilon^3 \sum_{\stackrel
{\scriptstyle n_{3}
= 1} {n_{3} \neq n_{1},n_{2}}}^{N} \left[
\frac{1}{|\boldsymbol{x}_{n_3}-\boldsymbol{x}_{n_2}|^{3}} +
\frac{\epsilon^3}{|\boldsymbol{x}_{n_3}-\boldsymbol{x}_{n_1}|^{3}
|\boldsymbol{x}_{n_2}-\boldsymbol{x}_{n_1}|^{3}} \right]. \end{equation}

The equation governing $\boldsymbol{f-f}_{1}^{(n)}$ may be obtained from

(\ref{int30}) and (\ref{av30}) as
\begin{eqnarray}
\label{A18}
\samepage
& \displaystyle {\frac{1}{2} (f_{i}
-f_{1i}^{(n_{1})})
(\boldsymbol{y},\boldsymbol{x,x}_{1},\ldots,\boldsymbol{x}_{N})
= \bar{f}_{i}(\boldsymbol{y}) +} & \nonumber \\ & \displaystyle{ +
\int_{\partial s_{n_1}} (\sigma_{y})_{ij}(\boldsymbol{T}_{.k}) (f_{k} -
f_{1k}^{(n_{1})})(\boldsymbol{x,x}_{1},\ldots,\boldsymbol{x}_{N})
ds_{x}\hat{n}_{j}(\boldsymbol{y})} + \nonumber \\ & \displaystyle{+
\sum_{\stackrel {\scriptstyle n_{2} = 1} {n_{2}
\neq n_{1}}}
^{N} \int_{\partial s_{n}}
(\sigma_{y})_{ij}(\boldsymbol{T}_{.k})
[f_{k}-f_{0k}^{(n_{1},n_{2})}(\boldsymbol{x}_{n_{1}},
\boldsymbol{x}_{n_2})] ds_{x} \hat{n}_{j}(\boldsymbol{y}).}
\end{eqnarray}
Consequently, upon utilizing (\ref{A17}) one obtains \begin{eqnarray}
\label{A19}
& \displaystyle{\| \boldsymbol{f-f}_{1}^{(n_{1})}
\|_{L^{\infty}(\partial
s_{n_1})} \leq C \mu \| \boldsymbol{G} \| \epsilon^{6}} \cdot
& \nonumber \\ & \displaystyle{ \cdot \sum_{\stackrel
{\scriptstyle n_{2}
= 1} {n_{2} \neq n_{1}}}^{N}
\sum_{\stackrel {\scriptstyle n_{3}
= 1} {n_{3} \neq n_{1},n_{2}}}^{N}
\left[
\frac{1}{|\boldsymbol{x}_{n_3}-\boldsymbol{x}_{n_2}|^{3}
|\boldsymbol{x}_{n_2}-\boldsymbol{x}_{n_1}|^{3}} +
\frac{\epsilon^3}{|\boldsymbol{x}_{n_3}-\boldsymbol{x}_{n_1}|^{3}
|\boldsymbol{x}_{n_2}-\boldsymbol{x}_{n_1}|^{6}} \right].} &
\end{eqnarray}
The error in the average velocity is therefore \begin{eqnarray}
\label{A20}
| \left< \boldsymbol{u} -\bar{\boldsymbol{u}} -
\boldsymbol{u}_{1} - \boldsymbol{u}_{2} \right> | & \leq & \epsilon^{9}
C
\| \boldsymbol{G} \| \int_{\Omega}
\frac{n(\boldsymbol{x}_{1})
dx_{1}}{|\boldsymbol{x}_{1}-\boldsymbol{x}|^{2}} \left[ \int_{\Omega}
\frac{P(\boldsymbol{x}_{2}/\boldsymbol{x}_{1}) dx_{2}}
{|\boldsymbol{x}_{2}-\boldsymbol{x}_{1}|^{3}} \int_{\Omega}
\frac{P(\boldsymbol{x}_{3}/\boldsymbol{x}_{2},\boldsymbol{x}_{1})
dx_{3}}
{|\boldsymbol{x}_{3}-\boldsymbol{x}_{2}|^{3}} +\right. \nonumber \\ & &
+
\left. \int_{\Omega}
\frac{\epsilon^{3}P(\boldsymbol{x}_{2}/\boldsymbol{x}_{1}) dx_{2}}
{|\boldsymbol{x}_{2}-\boldsymbol{x}_{1}|^{6}} \int_{\Omega}
\frac{P(\boldsymbol{x}_{3}/\boldsymbol{x}_{2},\boldsymbol{x}_{1})
dx_{3}}
{|\boldsymbol{x}_{3}-\boldsymbol{x}_{1}|^{3}} \right], \end{eqnarray}
which is valid in the limit $N \to \infty$ if (\ref{A05}) is
correct.

\newsec{The error in the average velocity due to wall effects and
spatial
inhomogeneties} Write (\ref{av10}) in the form
\begin{eqnarray}
\label{B1}
& & \frac{1}{2}f_{0i}^{n}(\boldsymbol{y}) = [2 \mu
G_{ij}(\boldsymbol{x}_{n}) -
\bar{p}(\boldsymbol{x}_{n})] \hat{n}_{j}(\boldsymbol{y}) + \frac{3}{4
\pi}
\int_{\partial s_{n}} \frac{r_{i} r_{j} r_{k}}{r^5}
f_{0k}^{(n)}(\boldsymbol{x,x}_{n})
ds_{x}\hat{n}_{j}(\boldsymbol{y})
+ \{ 2 \mu [G_{ij}(\boldsymbol{y}) - \nonumber \\ & & -
G_{ij}(\boldsymbol{x}_{n})] -
[\bar{p}(\boldsymbol{y}) - \bar{p}(\boldsymbol{x}_{n})] \}
\hat{n}_{j}(\boldsymbol{y}) + \int_{\partial s_{n}}
(\sigma_{y})_{ij}(\boldsymbol{\tau}_{.k}) (\boldsymbol{x,y})
f_{0k}^{(n)}(\boldsymbol{x},\boldsymbol{x}_{n})
ds_{x}\hat{n}_{j}(\boldsymbol{y}).
\end{eqnarray}
Obviously,
\begin{asubequation}
\label{B2}
\begin{eqnarray}
\| \boldsymbol{G}(\boldsymbol{y}) - \boldsymbol{G}(\boldsymbol{x}_{n})
\|_{L^{\infty}(\Omega)} \leq C \frac{\epsilon}{R}
\|\boldsymbol{G}\|_{L^{\infty}(\Omega)}, \\ \| \bar{p}(\boldsymbol{y}) -
\bar{p}(\boldsymbol{x}_{n}) \|_{L^{\infty}(\Omega)} \leq C
\frac{\epsilon}{R} \|\bar{p}\|_{L^{\infty}(\Omega)}
\end{eqnarray}
\end{asubequation}
\hspace*{-1.5em} and
\begin{equation}
\label{B3}
\left| \int_{\partial s_{n}}
(\sigma_{y})_{ij}(\boldsymbol{\tau}_{.k})(\boldsymbol{ x,y})
f_{0k}^{(n)}
ds_{x}\hat{n}_{j}(\boldsymbol{y}) \right| \leq C \epsilon^{3} \left|
\nabla_{x} \left.
\sigma_{y}(\boldsymbol{\tau})\right|_{(\boldsymbol{x}_{n},\boldsymbol{y})}
\right| \mu \|\boldsymbol{G}\|.
\end{equation}
Upon utilizing the fact that
$\boldsymbol{\tau(x,y)}$ is
uniformly bounded throughout the whole domain, except when both
$\boldsymbol{x}$ and $\boldsymbol{y}$ lie near \(\partial \Omega \), we
obtain the rather crude estimate \begin{displaymath}
\boldsymbol{\nabla}\boldsymbol{\tau}(\boldsymbol{x},\boldsymbol{y}) \leq
\frac{C}{d(\boldsymbol{x},\partial \Omega)^{3}} \; , \end{displaymath}
Hence,
\begin{equation}
\label{B4}
\left| \int_{\partial s_{n}}
(\sigma_{y})_{ij}(\boldsymbol{\tau}_{.k})(\boldsymbol{ x,y})
f_{0k}^{(n)}
ds_{x}\hat{n}_{j}(\boldsymbol{y}) \right| \leq C
\frac{\epsilon^{3}}{d(\boldsymbol{x}_{n},\partial
\Omega)^{3}} \mu \|\boldsymbol{G}\|.
\end{equation}
Thus,
\begin{equation}
\label{B5}
\| \boldsymbol{f}_{0}^{n} - 5 \mu \boldsymbol{G} \cdot
\hat{\boldsymbol{n}} +
\bar{p}\hat{\boldsymbol{n}} \|_{L^{\infty}(\partial s_{n})} \leq C\mu
\|\boldsymbol{G}\| \left[
\frac{\epsilon^{3}}{d(\boldsymbol{x}_{n},\partial
\Omega)^{3}} + \frac{\epsilon}{R} \right]. \end{equation}
The error in the configuration-dependent velocity field may now be
estimated as
\begin{equation}
\label{B6}
|\Delta \boldsymbol{u}_{1}| \leq C \|\boldsymbol{G}\| \sum_{n=1}^{N}
\frac{\epsilon^3}{|\boldsymbol{x-x}!_{n}|^{2}}\left[
\frac{\epsilon^{3}}{d(\boldsymbol{x}_{n},\partial \Omega)^{3}} +
\frac{\epsilon}{R} \right].
\end{equation}
Consequently, the error in the average velocity is \begin{equation}
\label{B7}
| \left< \Delta \boldsymbol{u}_{1} \right> | \leq C
\|\boldsymbol{G}\|
\left[ \int_{\Omega} \frac{\epsilon^{6} n(\boldsymbol{x}_{1})
dx_{1}}{|\boldsymbol{x-x}_{1}|^{2} d(\boldsymbol{x}_{1},\partial
\Omega)^{3}} +
\epsilon \bar{C} \right].
\end{equation}
Motivated by the fact that $d(\boldsymbol{x},\partial \Omega) \gg
\epsilon$, we estimate the first term in brackets as 
\begin{displaymath}
\begin{array}{rcl}
\displaystyle{\int_{\Omega} \frac{\epsilon^{6} n(\boldsymbol{x}_{1})
dx_{1}}{|\boldsymbol{x-x}_{1}|^{2} d(\boldsymbol{x}_{1},\partial
\Omega)^{3}}} & \leq &
\displaystyle{ \frac{8 \epsilon^3}{d(\boldsymbol{x},\partial
\Omega)^{3}}
\int_{B(\boldsymbol{x},\frac{1}{2}d(\boldsymbol{x},\partial
\Omega))}\frac{\epsilon^{3} n(\boldsymbol{x}_{1})
dx_{1}}{|\boldsymbol{x-x}_{1}|^{2}} +} \\ & & + \displaystyle{\frac{4
\epsilon^2}{d(\boldsymbol{x},\partial
\Omega)^{2}}
\int_{\Omega/B(\boldsymbol{x},\frac{1}{2}d(\boldsymbol{x},\partial
\Omega))}\frac{\epsilon^{3} n(\boldsymbol{x}_{1})
dx_{1}}{d(\boldsymbol{x}_{1},\partial \Omega)^{3}}} \end{array}
\end{displaymath}
or
\begin{displaymath}
\int_{\Omega} \frac{\epsilon^{6} n(\boldsymbol{x}_{1})
dx_{1}}{|\boldsymbol{x-x}_{1}|^{2} d(\boldsymbol{x}_{1},\partial
\Omega)^{3}} \leq c R
\left[ C_{1} \frac{\epsilon^3}{d(\boldsymbol{x},\partial
\Omega)^{3}} + C_{2} \frac{R \epsilon}{d(\boldsymbol{x},\partial
\Omega)^{2}} \right],
\end{displaymath}
whence
\begin{equation}
\label{B8}
| \left< \Delta \boldsymbol{u}_{1} \right> | \leq C
\|\boldsymbol{G}\| R c \;
\frac{R \epsilon}{d(\boldsymbol{x},\partial \Omega)^2}. \end{equation}
More refined estimates can probably be derived from (\ref{B7}); however,
for $d(\boldsymbol{x},\partial \Omega) \sim O(R)$ the error is of
$O(c\epsilon/R)$, which is sufficiently small to justify the degree of
omission made in utilizing (\ref{hom20}). 
\bibliography{work1}

\begin{thebibliography}{10}
\bibitem{lenote}
Details of the proof can be obtained directly from the authors.
\end{thebibliography}
\end{document}